\documentclass[12pt,preprint]{aastex}
\usepackage{color}

\begin{document}

\title{The serendipitous observation of a gravitationally lensed galaxy at z = 0.9057 from
the Blanco Cosmology Survey: The Elliot Arc}


\def\Munich{9}
\def\ExcellenceCluster{10}
\def\Michigan{18}
\def\Illinois{15}
\def\NCSA{2}
\def\MPE{11}
\def\CUTaiwan{14}
\def\UNDakota{3}
\def\Fermilab{1}
\def\IAPFrance{4}
\def\CfA{5}
\def\KECKfellow{6}
\def\UCOLick{8}
\def\UChicago{7}
\def\Tokyo{12}
\def\IAAASTaiwan{13}
\def\STSI{16}
\def\NOAO{17}
\def\Exeter{19}

\author{
E. J. Buckley-Geer,\altaffilmark{\Fermilab}
H. Lin,\altaffilmark{\Fermilab}
E. R. Drabek,  \altaffilmark{\Fermilab,\Exeter}
S. S. Allam,\altaffilmark{\Fermilab}
D. L. Tucker,\altaffilmark{\Fermilab}
R. Armstrong,\altaffilmark{\NCSA} 
W. A. Barkhouse,\altaffilmark{\UNDakota} 
E. Bertin,\altaffilmark{\IAPFrance}
M. Brodwin, \altaffilmark{\CfA,\KECKfellow}
S. Desai,\altaffilmark{\Illinois} 
J. A. Frieman,\altaffilmark{\Fermilab,\UChicago} 
S. M. Hansen,\altaffilmark{\UCOLick}
F. W. High,\altaffilmark{\UChicago}
J. J. Mohr,\altaffilmark{\Munich,\ExcellenceCluster,\MPE}
Y.-T. Lin,\altaffilmark{\Tokyo,\IAAASTaiwan}
C.-C. Ngeow,\altaffilmark{\CUTaiwan,\Illinois}
A. Rest,\altaffilmark{\STSI}
R. C. Smith,\altaffilmark{\NOAO} \\
J. Song,\altaffilmark{\Michigan}
A. Zenteno,\altaffilmark{\Munich,\ExcellenceCluster}
}

\altaffiltext{\Fermilab}{Center for Particle Astrophysics, Fermi National Accelerator Laboratory, P.O. Box 500, Batavia, IL 60510}
\altaffiltext{\NCSA}{National Center for Supercomputing Applications,University of Illinois, 1205 West Clark Street, Urbanan, IL 61801}
\altaffiltext{\UNDakota}{Department of Physics \& Astrophysics,University of North Dakota, Grand Forks, ND 58202}
\altaffiltext{\IAPFrance}{Institut d'Astrophysique de Paris, UMR 7095 CNRS, Universit\'e Pierre et Marie Curie, 98 bis boulevard Arago, F-75014 Paris, France}
\altaffiltext{\CfA}{Harvard-Smithsonian Center for Astrophysics, 60 Garden Street, Cambridge, MA 02138}
\altaffiltext{\KECKfellow}{W. M. Keck Postdoctoral Fellow at the Harvard-Smithsonian Center for Astrophysics}
\altaffiltext{\UChicago}{Department of Astronomy and Astrophysics, University of Chicago, 5640 South Ellis Avenue, Chicago, IL 60637}
\altaffiltext{\UCOLick}{National Science Foundation Astronomy \& Astrophysics Postdoctoral Fellow, University of California Observatories \& Department of Astronomy, University of California, Santa Cruz, CA 95064}
\altaffiltext{\Munich}{Department of Physics, Ludwig-Maximilians-Universit\"{a}t, Scheinerstr.\ 1, 81679 M\"{u}nchen, Germany}
\altaffiltext{\ExcellenceCluster}{Excellence Cluster Universe, Boltzmannstr.\ 2, 85748 Garching, Germany}
\altaffiltext{\MPE}{Max-Planck-Institut f\"{u}r extraterrestrische Physik, Giessenbachstr.\ 85748 Garching, Germany}
\altaffiltext{\Tokyo}{Institute for Physics and Mathematics of the Universe, University of Tokyo, 5-1-5 Kashiwa-no-ha, Kashiwa-shi, Chiba 277- 8568, Japan}
\altaffiltext{\IAAASTaiwan}{Institute of Astronomy \& Astrophysics, Academia Sinica, Taipei, Taiwan}
\altaffiltext{\CUTaiwan}{Graduate Institute of Astronomy, National Central University, No. 300 Jonghda Rd, Jhongli City 32001 Taiwan} 
\altaffiltext{\Illinois}{Department of Astronomy, University of Illinois, 1002 West Green Street, Urbana, IL 61801}
\altaffiltext{\STSI}{Space Telescope Science Institute, 3700 San Martin Dr., Baltimore, MD 21218}
\altaffiltext{\NOAO}{Cerro Tololo Inter-American Observatory, National
Optical Astronomy Observatory, La Serena, Chile}
\altaffiltext{\Michigan}{Department of Physics, University of Michigan, 450 Church St. Ann Arbor, MI 48109}
\altaffiltext{\Exeter}{School of Physics, University of Exeter, Stocker Road, Exeter EX4 4QL, United Kingdom}






\begin{abstract}
We report on the serendipitous discovery in the Blanco Cosmology
Survey (BCS) imaging data of a $z = 0.9057$
galaxy that is being strongly lensed by a massive galaxy cluster
at a redshift of $z=0.3838$. 
The lens (BCS J2352-5452) was discovered while examining $i$- and $z$-band
images being acquired in October 2006 during a BCS observing run.
Follow-up spectroscopic observations with the GMOS instrument on the
Gemini South 8m telescope confirmed the lensing nature of
this system.
Using weak plus strong lensing, velocity dispersion, cluster richness
$N_{200}$, and fitting to an NFW cluster mass density profile,
we have made three independent estimates of the mass $M_{200}$
which are all very consistent with each other. 
The combination of
the results from the three methods gives $M_{200} = (5.1 \pm 1.3)
\times 10^{14} M_\Sun$, which is fully consistent with the individual
measurements.  The final NFW concentration $c_{200}$
from the combined fit is $c_{200} = 5.4^{+1.4}_{-1.1}$. 
We have compared our measurements of $M_{200}$ and $c_{200}$ with
predictions for (a) clusters from $\Lambda$CDM simulations, (b)
lensing selected clusters from simulations, and (c) a real sample of
cluster lenses.  We find that we are most
compatible with the predictions for $\Lambda$CDM simulations for lensing 
clusters, and we see no evidence based on this one system for an increased
concentration compared to $\Lambda$CDM.
Finally, using the flux
measured from the [OII]3727 line  we have determined
the star formation rate (SFR) of the source galaxy and find it to be rather modest given the
assumed lens magnification.

\end{abstract}





\section{Introduction}
Strong gravitational lenses offer unique opportunities to study cosmology, dark matter, galactic
structure, and galaxy evolution. They also provide a sample of galaxies, namely the lenses themselves, that are
selected based on total mass rather than luminosity or surface brightness. The majority of lenses
discovered in the past decade were found through dedicated surveys using a variety of techniques.
For example, the Sloan Digital Sky Survey (SDSS) data have been used to effectively select lens
candidates from rich clusters \citep{hennawi08} through intermediate
scale clusters \citep{allam07,lin09} to
individual galaxies \citep{bolton08,willis06}. Other
searches using the CFHTLS \citep{cabanac07} and COSMOS fields
\citep{faure08,jackson08} have yielded 40 and 70 lens
candidates respectively. These
searches cover the range of 
giant arcs with Einstein radii $\theta_{EIN}> 10\arcsec$ all the way to
small arcs produced by single lens galaxies with $\theta_{EIN}<
3\arcsec$. 

In this paper we report on the serendipitous discovery of a strongly
lensed $z=0.9057$ galaxy in the Blanco Cosmology Survey (BCS) imaging
data. The lens is a rich cluster containing a prominent central 
brightest cluster galaxy (BCG) and has a redshift of $z=0.3838$. 
Cluster-scale lenses are particularly useful as they allow
us to study the effects of strong lensing in the core of the cluster and
weak lensing in the outer regions. Strong lensing provides constraints
on the mass contained within the Einstein radius of the arcs whereas weak
lensing provides information on the mass profiles in the outer reaches
of the cluster.  Combining the two measurements allows us to make
tighter constraints on the mass $M_{200}$ and the concentration $c_{200}$,
of an NFW \citep{nfw95} model of the cluster mass density profile,
over a wider range of radii than would be possible with
either method alone
\citep{natarajan98,natarajan02,bradac06,bradac08a,bradac08b,diego07,limousin07,hicks07,deb08,merten09,oguri09}.
In addition, if one has spectroscopic redshifts for the
member galaxies one can determine the cluster velocity dispersion, assuming the cluster
is virialized, and hence obtain an independent estimate for $M_{200}$
\citep{becker07}. Finally one can also derive an $M_{200}$ estimate
from the maxBCG cluster richness $N_{200}$
\citep{hansen05,johnston07}.  These three different methods, strong plus weak
lensing, cluster velocity dispersion, and optical richness, provide independent
estimates of $M_{200}$ ($M_{200}$ is defined as the mass within a sphere of overdensity 200
times the critical density at the redshift $z$) and can then be combined to obtain improved
constraints on $M_{200}$ and $c_{200}$.  Measurements of the concentration
from strong lensing clusters is of particular interest as recent
publications suggest that they may be
more concentrated than one would expect from $\Lambda$CDM models
\citep{broadhurst_barkana08,oguri_blandford09}.

The paper is organized as follows. In \S~\ref{sec:bcs_survey} we describe 
the Blanco Cosmology Survey. Then in \S~\ref{sec:discovery} we discuss the
initial discovery and the spectroscopic follow-up that led to confirmation of the system as a
gravitational lens, the data reduction, the properties of the cluster,
the extraction of the redshifts, and finally the measurement of the
cluster velocity dispersion and estimate of the cluster mass.
In \S~\ref{sec:lens_modeling} we summarize the strong
lensing features of the system. In 
\S~\ref{sec:weak_lens} we describe the weak lensing measurements. 
In \S~\ref{sec:combined_constraints} we present the results of combining of the strong
and weak lensing results and the final mass
constraints derived from combining the lensing results with the velocity
dispersion and richness measurements. 
We describe the source galaxy star formation rate
measurements in \S~\ref{sec:SFR} and finally in \S~\ref{sec:conclusions} we conclude. 
We assume a flat cosmology with $\Omega_{\rm M}=  0.3$,
$\Omega_{\Lambda}=0.7$, and
$H_0=70$~km~s$^{-1}$~Mpc$^{-1}$, unless otherwise noted.

\section{The BCS Survey}\label{sec:bcs_survey}

The Blanco Cosmology Survey (BCS) is a 60-night NOAO
imaging survey program (2005-2008), using the Mosaic-II camera on the 
Blanco 4m telescope at CTIO, that has
uniformly imaged $75 \deg^2$ of the sky in the SDSS $griz$ bands
in preparation for cluster finding with the South Pole Telescope (SPT)
\citep{vanderlinde10} and 
other millimeter-wave experiments. The depths in each band
were chosen to allow the estimation of photometric redshifts for $L\ge
L_*$ galaxies out to a redshift of $z = 1$ and to detect galaxies to
$0.5L_*$ at $5\sigma$ to these same redshifts. The survey was divided
into two fields to allow efficient use of the allotted nights between
October and December. Both fields lie near $\delta = -55^{\circ}$ which
allows for overlap with the SPT. One field is centered near
$\alpha=23.5$ hr and the other is at $\alpha=5.5$ hr. In addition to the large science fields,
BCS also covers 7 small fields that overlap large spectroscopic
surveys so that photometric redshifts (photo-z's) using
BCS data can be trained and tested using a sample
of over 5,000 galaxies.

\section{Discovery of the lens and spectroscopic follow-up}\label{sec:discovery}
\begin{figure}
\plotone{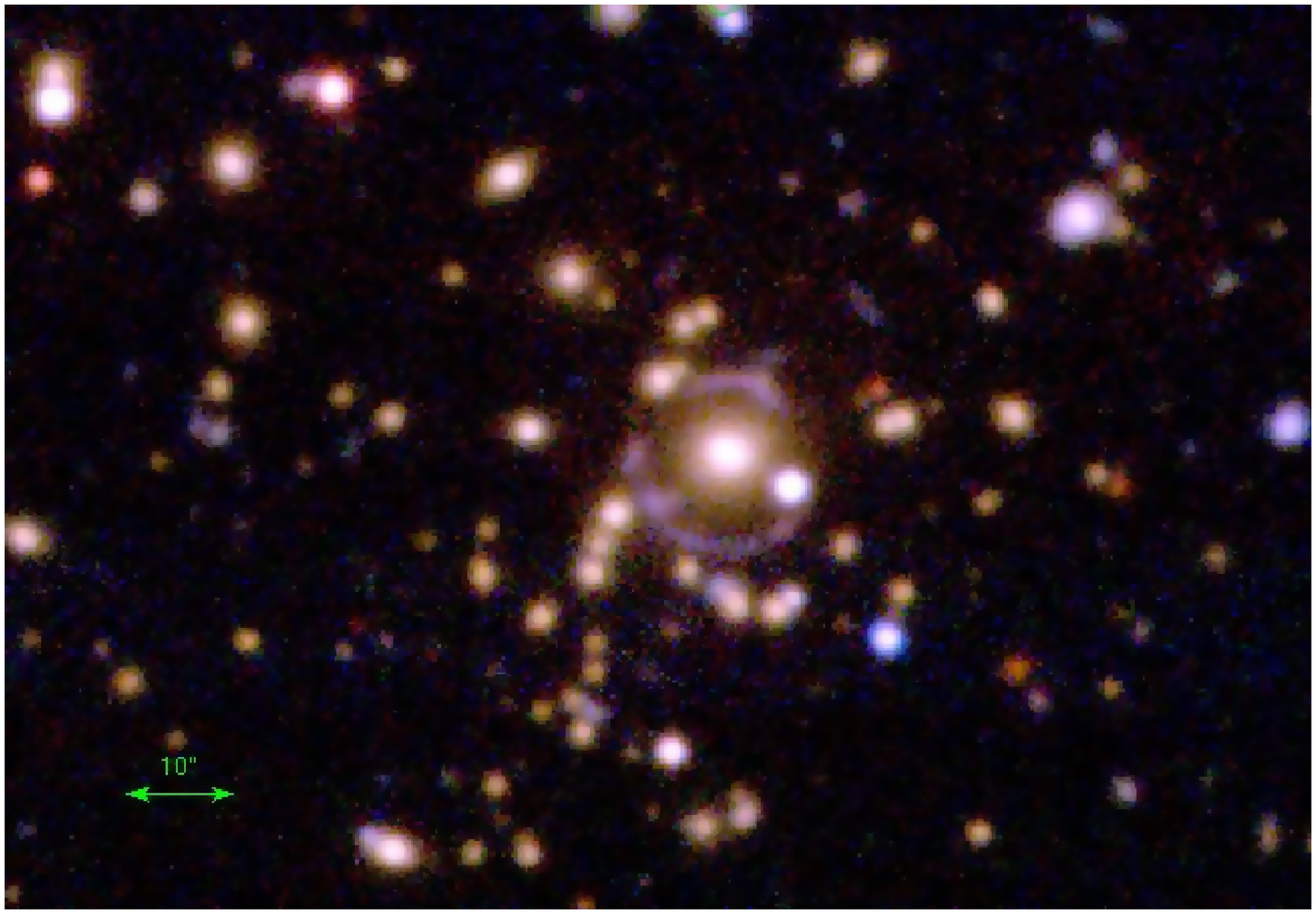}
\caption{A $gri$ color image of the Elliot Arc and its cluster
environment. The scale is indicated by the horizontal arrow. \label{color_coadd}}
\end{figure}

The lens BCS J2351-5452 was discovered serendipitously while examining $i$- and $z$-band
images being
acquired in October 2006 during the yearly BCS observing run. 
The discoverer (EJB-G) decided
to name it ``The Elliot Arc'' in honor of her then eight-year old
nephew.  Table~\ref{table_obslog} lists the observed images along
with seeing conditions. Fig.~\ref{color_coadd} shows a $gri$ color image
of the source, lens and surrounding environment (the pixel
scale is $0.268\arcsec$ per pixel). The source forms 
a purple ring-like structure of radius $\sim 7.5\arcsec$ with multiple distinct bright regions. The lens is
the BCG at the center of a large galaxy cluster. Photometric
measurements estimated the redshift of the cluster at $z\sim 0.4$, 
using the expected $g-r$ and $r-i$ red sequence colors,
and also provided a photo-z for the source of $z\sim 0.7$, 
as described below. We note that this cluster was first reported as SCSO~J235138$-$545253 in an independent analysis of the BCS data by \citet{menanteau10} where its remarkable lens was noted and they estimated a photometric redshift of $z=0.33$ for the cluster.
 
\begin{figure}
\plotone{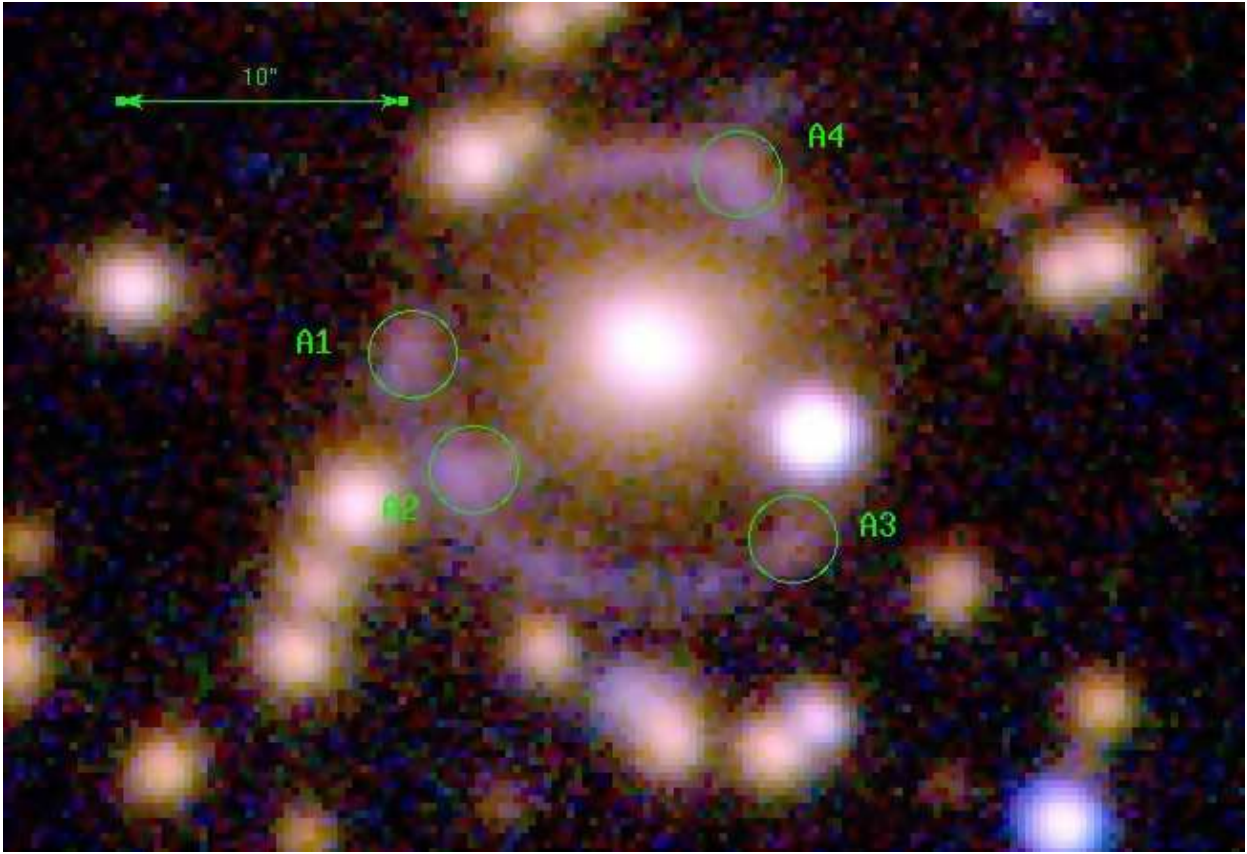}
\caption{A $gri$ color image of the Elliot Arc. The knots
targeted for spectroscopy are shown as green circles. The
scale is indicated by the horizontal line.\label{knot_targets}}
\end{figure}

\begin{figure}
\includegraphics[scale=0.75]{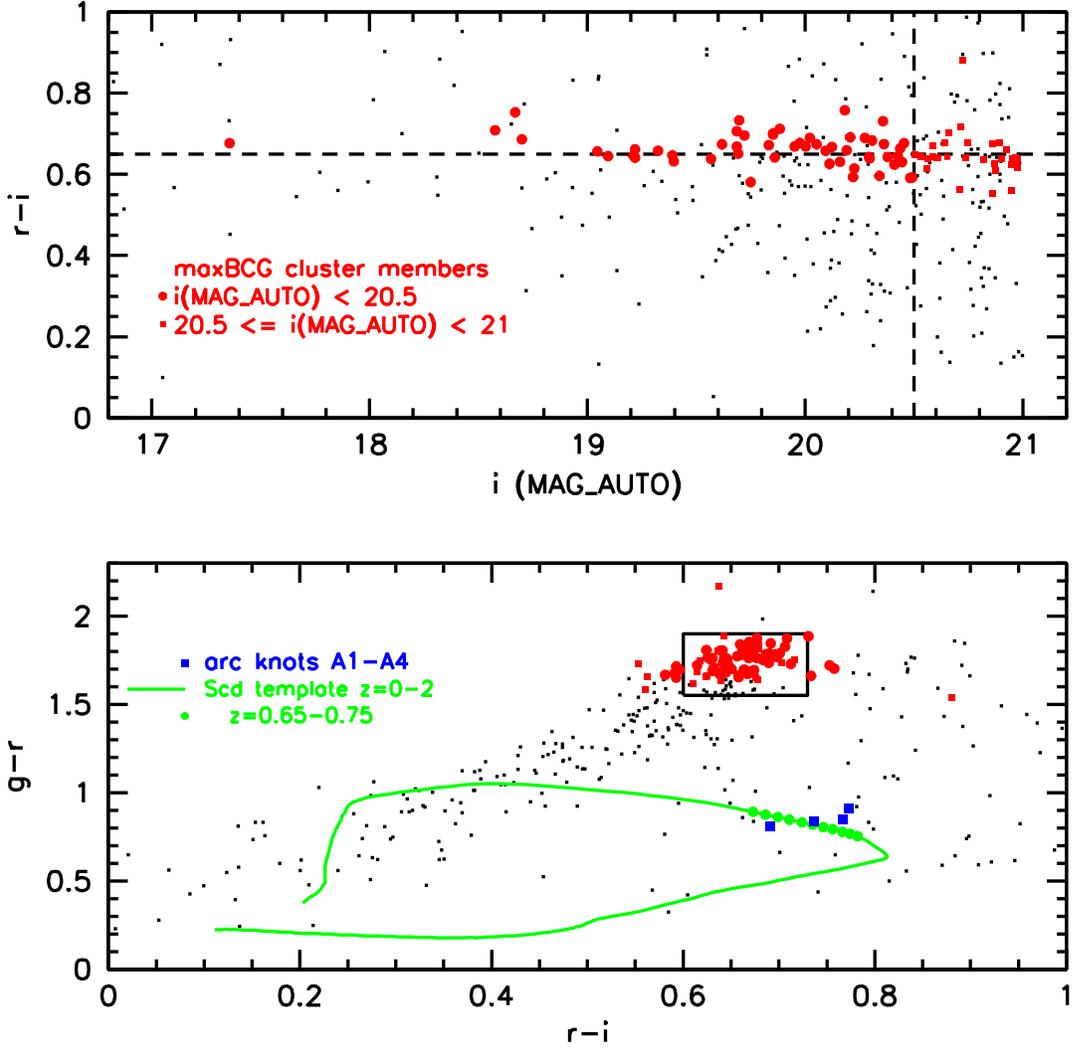}
\caption{(Top) $r-i$ vs.\ $i$({\tt MAG\_AUTO}) color-magnitude diagram for
all galaxies (black points) with $i < 21$ and within a radius
$r_{200}^{gal} = 1.51~h^{-1}~{\rm Mpc}$ ($= 6.88\arcmin$) of the BCG.
Colors are measured using $3\arcsec$-diameter aperture magnitudes.
Galaxies meeting the maxBCG cluster color selection criteria
(see \S\ref{sec:cluster_properties}) are plotted in red, with red
circles indicating cluster members brighter than $i = 20.5$,
and red squares indicating fainter cluster members.
(Bottom) $g-r$ vs. $r-i$ color-color diagram for the same galaxies as in
the top panel.  Red circles and squares again indicate brighter and fainter
maxBCG cluster members, while the black rectangle indicates the color
selection box (approximating the more detailed maxBCG color criteria)
used to select likely cluster galaxies for GMOS spectroscopy
(see \S\ref{sec:discovery}).
In addition, the 4 bright knots A1-A4 (Fig.~\ref{knot_targets}) in the lensed
arcs are shown by the blue squares.  The green curve is an Scd galaxy model
\citep{cww} at redshifts $z = 0-2$, with green circles highlighting
the redshift range $z = 0.65-0.75$, indicating an approximate photometric
redshift $z \sim 0.7$ for the arc knots.
\label{color-color}}
\end{figure}

We obtained Gemini Multi-Object Spectrograph (GMOS)
spectra of the source and a number of the neighboring galaxies \citep{lin07}.
We targeted the regions of the source
labeled A1-A4 in Fig.~\ref{knot_targets}, and photometric properties of these
bright knots are summarized in Table~\ref{table_knots}.  
In addition we selected 51 more objects for a total of 55 spectra.  
The additional objects were selected using
their colors in order to pick out likely cluster member galaxies.
Fig.~\ref{color-color} shows the $r-i$ versus $i$ color-magnitude diagram
(top plot) and the $g-r$ vs.\ $r-i$ color-color diagram (bottom plot) of
the field.
The blue squares in the bottom panel of Fig.~\ref{color-color} show the
four targeted knots in the lensed arcs.
The green curve is an Scd galaxy model \citep{cww} with the green circles
indicating a photometric redshift for the arc of $z \sim 0.7$.
Note this is not a detailed photo-z fit, but is just a rough estimate meant
to show that the arc is likely at a redshift higher than the cluster redshift.
Highest target priority was given to the arc knots and to the BCG.
Then cluster red sequence galaxy targets were selected using the simple
color cuts $1.55 \leq g-r \leq 1.9$ and $0.6 \leq r-i \leq 0.73$
(also shown in the bottom panel of Fig.~\ref{color-color}),
which approximate the more detailed final cluster membership criteria
described below in \S\ref{sec:cluster_properties}.
Red sequence galaxies with $i < 21.6$ ($3\arcsec$-diameter SExtractor aperture
magnitudes) were selected, with higher priority given to brighter galaxies
with $i(3\arcsec) \leq 21$.  Additional non-cluster targets lying
outside the cluster color selection box were added at lowest priority.

We used the GMOS R150 grating + the GG455 filter in order
obtain spectra with about 4600 -- 9000 \AA \ wavelength coverage.
This was designed to cover the [OII] 3727 emission line
expected at $\sim 6300$ \AA, given the photo-z estimate of $\sim 0.7$ for the arcs
as well as the Mg absorption features at $\sim 7000$ \AA \ (and the
4000 \AA \ break at $\sim 5600$ \AA) for the $z \sim 0.4$ cluster
elliptical galaxies.

We used 2 MOS masks in order to fully
target these cluster galaxies (along with the arcs) for spectroscopy.
Each mask had a 3600 second
exposure time split into 4 900-second exposures for cosmic ray
removal. We also took standard Cu-Ar lamp spectra for wavelength
calibrations and standard star spectra for flux calibrations. All data
were taken in queue observing mode. A summary of the observations is
given in Table~\ref{table_obslog}.

\subsection{Data Reduction}\label{sec:reduction}

The BCS imaging data were processed using the Dark Energy Survey data
management system (DESDM V3) which is under development at UIUC/NCSA/Fermilab
\citep{mohr08,ngeow06,zenteno11}. 
The images are corrected for instrumental effects which include crosstalk
correction, pupil ghost correction, overscan correction, trimming,
bias subtraction, flat fielding and illumination correction.
The images are then astrometrically calibrated and
remapped for later coaddition. For
photometric data, a photometric calibration is applied to the single-epoch 
and coadd object photometry.
The Astr$\cal O$matic software\footnote{http://www.astromatic.net}
SExtractor \citep{bertin96}, SCAMP \citep{bertin06} and SWarp \citep{bertin02} are used for
cataloging, astrometric refinement and remapping for coaddition over each image.
We have used the coadded images in the $griz$ bands for this
analysis. 

The spectroscopic data were processed using the standard data
reduction package provided by Gemini that runs in the IRAF
framework\footnote{http://www.gemini.edu/sciops/data-and-results/processing-software}.
We used version 1.9.1. This produced flux- and wavelength-calibrated 1-D spectra
for all the objects. Additional processing for the source spectra was done using the IRAF task
{\tt apall}.

\subsection {Cluster properties}\label{sec:cluster_properties}
We adopt the procedure used by the maxBCG cluster finder
\citep{koester07a,koester07b} to determine cluster membership and cluster
richness and to derive a richness-based cluster mass estimate.
We first measure $N_{gal}$, the number of cluster red sequence galaxies,
within a radius $1~h^{-1}$~Mpc ($= 4.55\arcmin$) of the BCG,
that are also brighter than $0.4 L_*$ at the cluster redshift $z = 0.38$.  
From \cite{koester07a}, $0.4 L_*$ corresponds to an $i$-band absolute 
magnitude $M = -20.25 + 5 \log h$ at $z = 0$, while at $z = 0.38$,
$0.4 L_*$ corresponds to an apparent magnitude $i = 20.5$ 
(specific value provided by J.\ Annis \& J.\ Kubo, private communication),
after accounting for both K-correction and evolution 
\citep[also as described in][]{koester07a}.  We apply this magnitude cut using
the SExtractor $i$-band {\tt MAG\_AUTO} magnitude, which provides a
measure of a galaxy's total light.  (Note the $3\arcsec$-diameter
aperture magnitude used earlier for target selection in general measures
less light cf.\ {\tt MAG\_AUTO}, but is better suited
for roughly approximating the light entering a GMOS slit.)
We set the red sequence membership cuts to be $g-r$ and $r-i$ color both
within $2\sigma$ of their respective central values $(g-r)_0 = 1.77$ and
$(r-i)_0 = 0.65$, where the latter are determined empirically based on the
peaks of the color histograms of galaxies within $1~h^{-1}$~Mpc of the BCG.
In applying the color cuts we use the colors defined by SExtractor
$3\arcsec$-diameter aperture magnitudes (this provides higher S/N colors
compared to using {\tt MAG\_AUTO}), and for the uncertainty
we define $\sigma = \sqrt{\sigma_{color}^2 + \sigma_{intrinsic}^2}$,
where $\sigma_{color}$ is the color measurement error derived from the
SExtractor aperture magnitude errors, and $\sigma_{intrinsic}$ is
the intrinsic red sequence color width, taken to be 0.05 for $g-r$
and 0.06 for $r-i$ \citep{koester07a}.

Carrying out the above magnitude and color cuts, we obtain an initial
richness estimate $N_{gal} = 44$.
Then, as discussed in \cite{hansen05}, we define another radius
$r_{200}^{gal} = 0.156~N_{gal}^{0.6}~h^{-1}~{\rm Mpc} = 1.51~h^{-1}~{\rm Mpc}$
($= 6.88\arcmin$), and repeat the same cuts within $r_{200}^{gal}$ of the BCG
to obtain a final richness estimate $N_{200} = 55$.
Finally, using the weak lensing mass calibration of \cite{johnston07}
for maxBCG clusters, we obtain a mass estimate
$M_{200} = (8.794 \times 10^{13}) \times (N_{200}/20)^{1.28}~h^{-1}~M_\Sun =
(4.6 \pm 2.1) \times 10^{14}~M_\Sun \ (h = 0.7)$,
where we have also adopted the fractional error of 0.45 derived
by \cite{rozo09} for this $N_{200}$-based estimate of $M_{200}$
for maxBCG clusters.

We note that \cite{rozo10} apply a factor of 1.18 to correct the
\cite{johnston07} cluster masses upward, in order to account for a 
photo-z bias effect that is detailed in \cite{mandelbaum08}.
We have not applied this correction as it makes only a 0.4$\sigma$ 
difference, although we remark that the resulting mass 
$M_{200} = 5.4 \times 10^{14}~M_\Sun$ does appear to improve the
(already good) agreement with our other mass estimates below 
(see \S\ref{sec:veldisp} and \S\ref{sec:combined}).

Fig.~\ref{color-color} shows color-magnitude and color-color plots of all
galaxies that have $i < 21$ (SExtractor {\tt MAG\_AUTO}) and that are
within a radius $r_{200}^{gal} = 1.51~h^{-1}~{\rm Mpc}$ ($= 6.88\arcmin$)
of the BCG.  Note we have extended the magnitude limit here down to
$i = 21$, to match the effective magnitude limit of our
spectroscopic redshift sample (\S\ref{sec:redshifts} below)
In particular, we find 86 maxBCG cluster members for $i < 21$, compared to
the earlier $N_{200} = 55$ for $i < 20.5$ (corresponding to $0.4 L_*$).
These member galaxies are shown using red symbols in Fig.~\ref{color-color}
and their properties are given in Table~\ref{cluster_galaxies}.

\subsection {Redshift determinations}\label{sec:redshifts}
\begin{figure}
\plotone{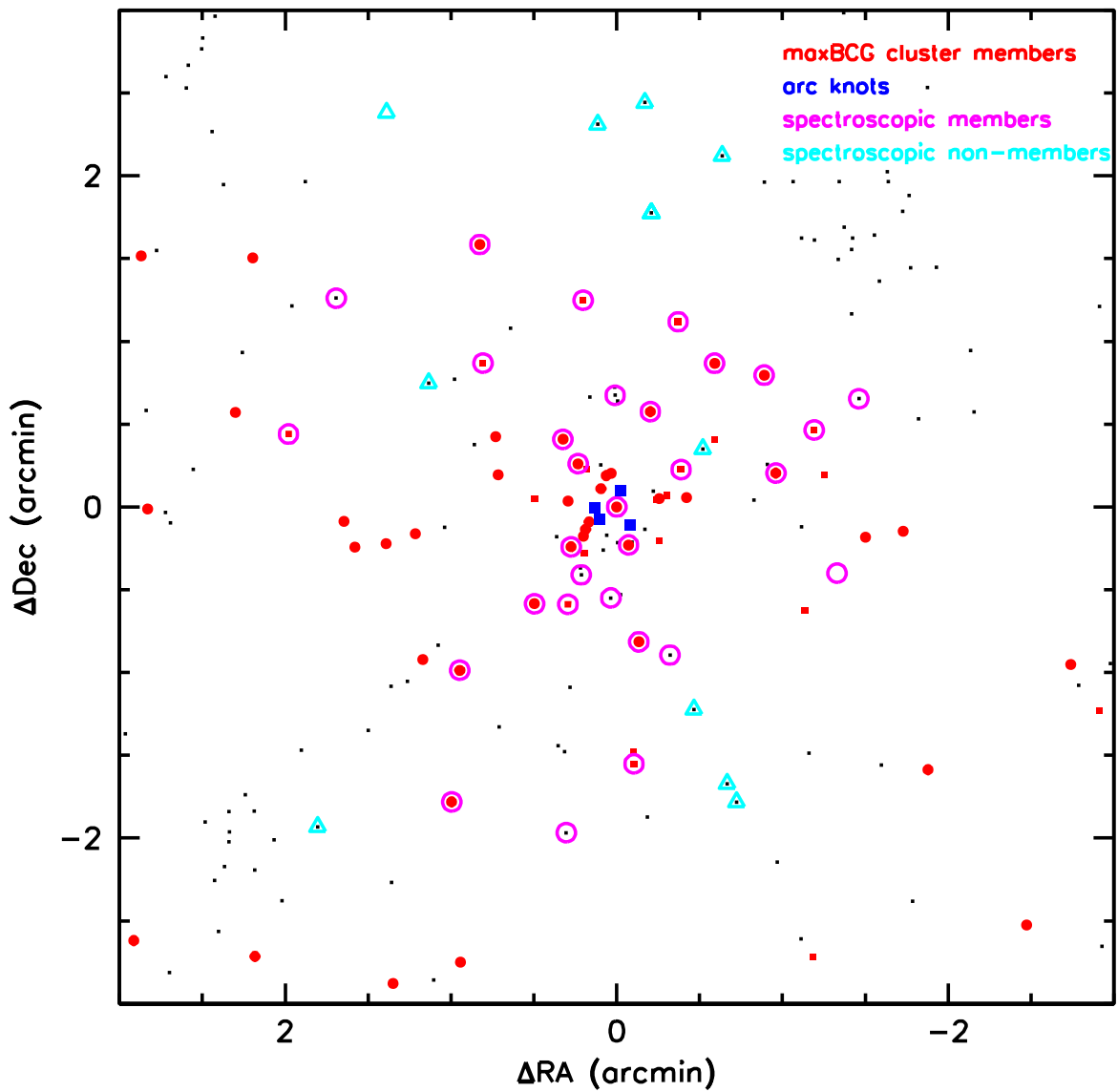}
\caption{Relative positions of all galaxies (points) with 
$i({\rm \tt MAG\_AUTO}) < 21$ within a $6\arcmin \times 6\arcmin$ 
box centered on the BCG. Cluster member galaxies defined using maxBCG criteria 
(see \S\ref{sec:cluster_properties}) are plotted in red, with red circles 
indicating members brighter than $i = 20.5$, and red squares indicating 
fainter members.  The 4 bright knots A1-A4 (Fig.~\ref{knot_targets}) in the 
lensed arcs are shown by the blue squares.  Galaxies determined to
be cluster members from GMOS redshifts are plotted with open magenta circles,
while those found spectroscopically to be non-members are shown with open 
cyan triangles (see \S\ref{sec:redshifts}).  North is up and East is to the
left.\label{all_targets}}
\end{figure}

\begin{figure}
\plotone{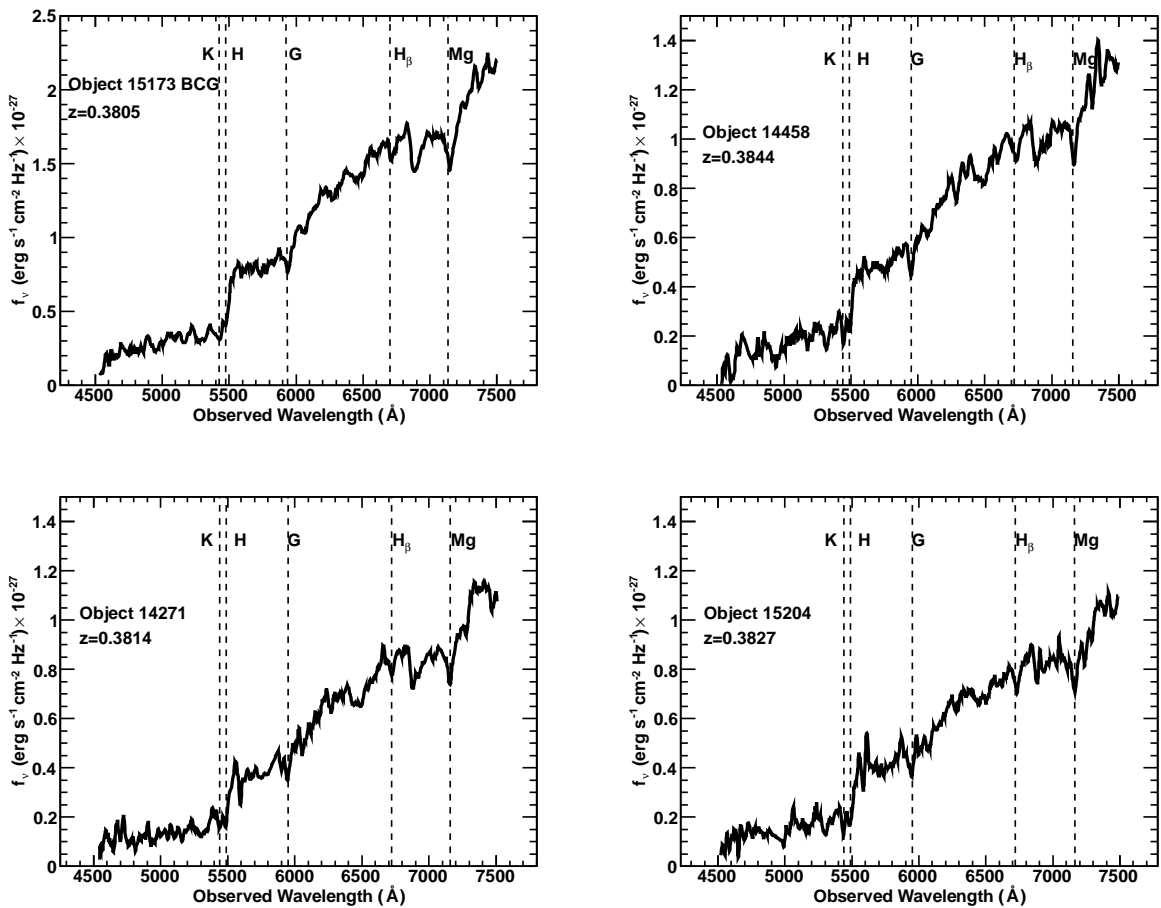}
\caption{Four examples of flux-calibrated cluster member spectra (in
$f_\nu$). The spectra have been smoothed (with a boxcar of 5 pixels =
17.8 \AA) to improve the signal-to-noise ratio. The spectrum in the top
left is that of the BCG. The prominent absorption features used in the
redshift identification are marked.\label{cluster_spectra}}
\end{figure}

The redshift extraction was carried out using the {\tt xcsao}
and {\tt emsao} routines in the IRAF external package {\tt rvsao}
\citep{kurtz98}. We obtained spectra for the 55 objects that were targeted.
Four of these spectra were of the source.  Out of the remaining 51 spectra we
had sufficient signal-to-noise in 42 of them to determine a redshift.
Thirty of the objects with redshifts between 0.377 and 0.393
constitute our spectroscopic sample of cluster galaxies.
Fig.~\ref{all_targets} shows the spatial distribution of galaxies within
a $6\arcmin \times 6\arcmin$ box centered on the BCG, with maxBCG cluster
members, arc knots, and objects with spectroscopic redshifts indicated
by different colors and symbols.
Table~\ref{cluster_galaxies} summarizes the properties of the 30 cluster
member galaxies with redshifts, and Table~\ref{other_galaxies} summarizes the
properties of the remaining 12 spectroscopic non-member galaxies.
In Fig.~\ref{cluster_spectra} we show four examples of the flux-calibrated
cluster member spectra including the BCG.

Examination of Table~\ref{cluster_galaxies} and Table~\ref{other_galaxies}
shows that our spectroscopic sample is effectively limited at
$i \approx 21$, as 39 of the 42 non-arc redshifts have $i < 21$.
Note that of the 30 spectroscopically defined cluster members, 22 are
also maxBCG members, while another 7 lie close to the maxBCG color selection
boundaries.  Also, of the 12 spectroscopic non-members, none
meets the maxBCG criteria except the faintest one (with $i = 21.58$).

\begin{figure}
\plotone{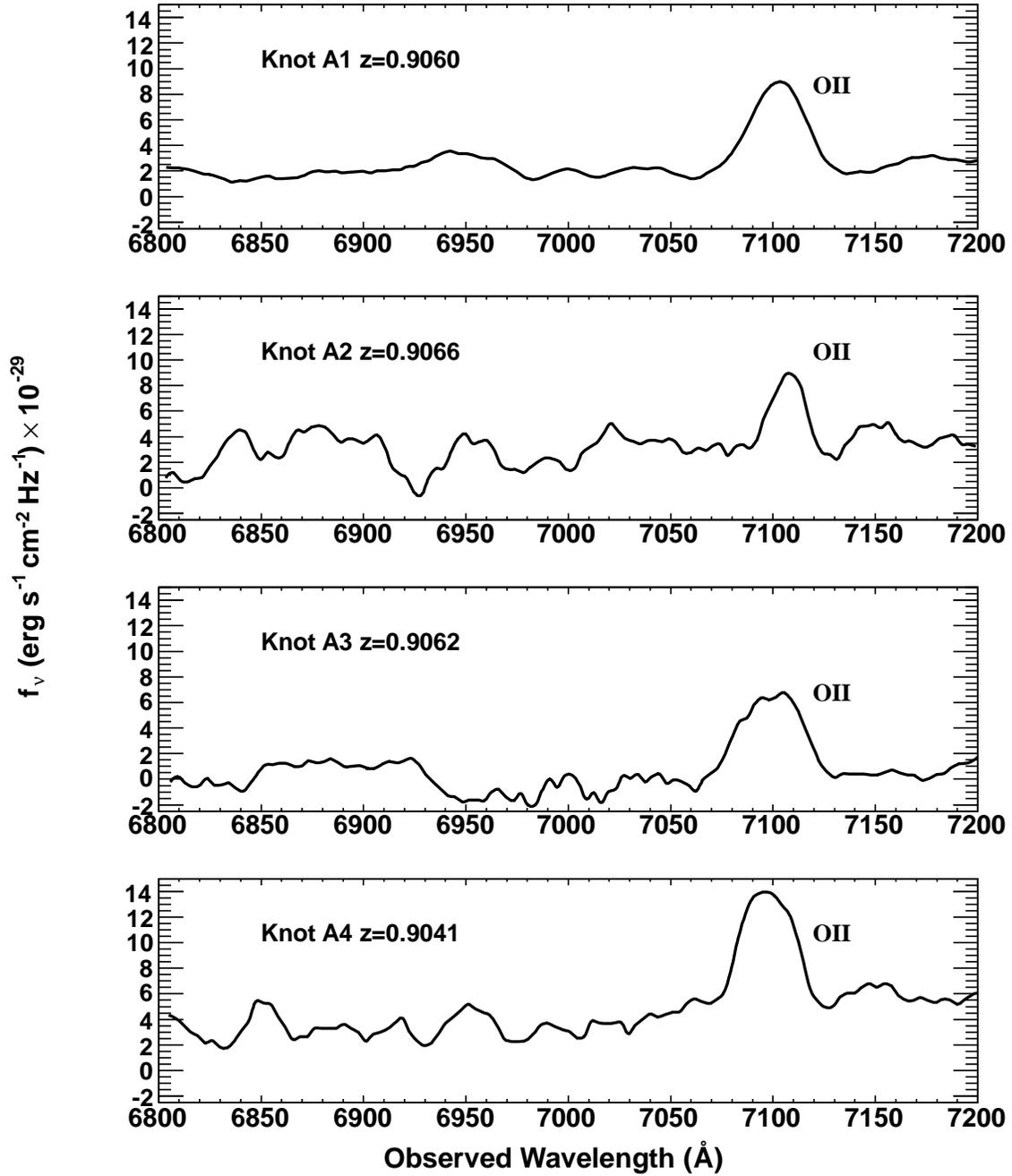}
\caption{Flux-calibrated spectra (in $f_\nu$) for the knots A1-A4.  The
spectra have been
smoothed (with a boxcar of 5 pixels = 17.8\AA) to improve S/N. Knot A2
was observed under seeing conditions that were a factor of two worse
than for the other three knots.
The [O II] 3727 \AA \ line is marked.
\label{arc_spectra}}
\end{figure}

The redshift of the source was determined from a single emission line
at 7100\AA \ which is present with varying signal-to-noise in each of the
knots that were observed. We take this line to be the [OII]3727\AA \ line which
yields a redshift of $0.9057\pm 0.0005$. The four
flux-calibrated source spectra are shown in Fig.~\ref{arc_spectra}. Knot A2
was observed under seeing conditions that were a factor of two worse
than for the other three knots (see Table~\ref{table_obslog}).

\subsection{Velocity dispersion and cluster mass measurement}\label{sec:veldisp}
\begin{figure}
\plotone{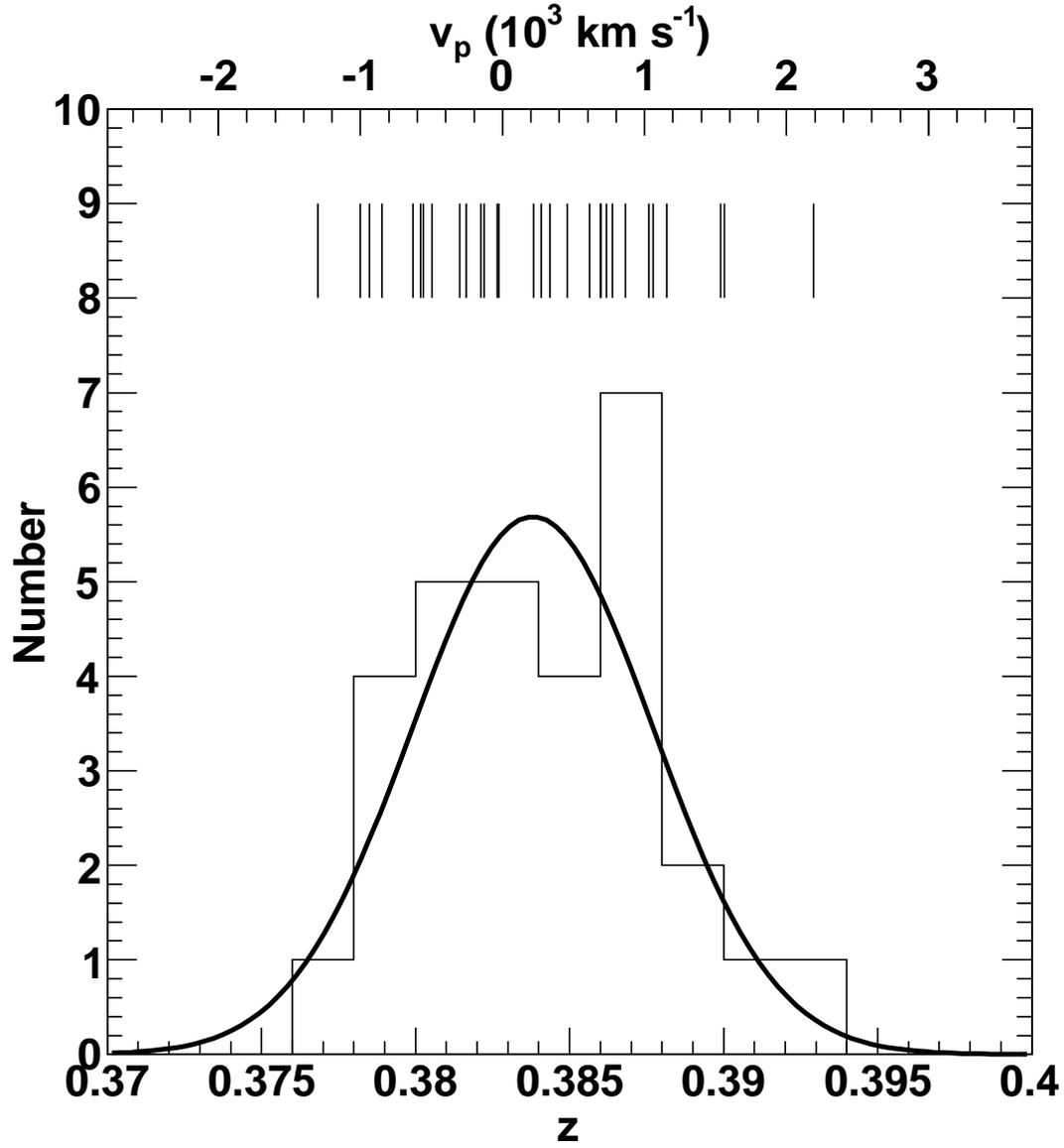}
\caption{The redshift distribution for the 30 cluster members in 
Table~\ref{cluster_galaxies}. The tick marks at the top represent the
individual cluster member peculiar velocities. The solid line is a Gaussian with mean
and sigma equal to $z_c$ and $\sigma_c \times (1+z_c)$ respectively
(see \S\ref{sec:veldisp}).\label{vel_disp}}
\end{figure}

We used the 30 cluster galaxies to estimate the redshift and velocity
dispersion of the cluster using the biweight estimators of
\citet{beers90}. We first use the biweight location estimator to determine the best estimate for {\it
cz}. This yields a value of $cz = 115151.1 \pm 241.1 \ {\rm km \ s^{-1}}$
which translates to a redshift of $z_c = 0.3838\pm 0.0008$. We then
use this estimate of the cluster redshift to determine the peculiar
velocity $v_p$ for each cluster member relative to the cluster center
of mass using 
\begin{equation}
v_p = \frac{(cz - cz_c)}{(1+z_c)} \label{eqn_vp}
\end{equation}
We determine the biweight estimate of scale for $v_p$ which is equal
to the velocity dispersion of the cluster. We find a value for the
velocity dispersion of $\sigma_c = 855^{+108}_{-96} \ {\rm km \
s^{-1}}$. The uncertainties are obtained by doing a jackknife
resampling. The
redshift distribution is shown in Fig.~\ref{vel_disp}. The overlaid
Gaussian has a mean of $z_c$ and a width of $\sigma_c\times (1+z_c)$. The lines
represent the individual peculiar velocities $v_p$ of the cluster members.

We can use the estimated velocity dispersion to derive an estimate for
the cluster mass. We use the results of \cite{evrard08} \citep[see also][]{becker07} which relates
$M_{200}$  to the dark matter velocity dispersion
\begin{equation}
M_{200} = 10^{15}\ M_{\sun} \frac{1}{h(z)}
\left(\frac{\sigma_{DM}}{\sigma_{15}} \right) ^{1/\alpha} \ , \label{eqn_m1}
\end{equation} 
where $h(z) = H(z)/100\ {\rm km\ s^{-1} Mpc^{-1}}$ is the dimensionless
Hubble parameter. The values of the parameters were found to be
$\sigma_{15} = 1082.9\pm 4\ {\rm km \ s^{-1}}$ and $\alpha = 0.3361\pm
0.0026$ \citep{evrard08}. Using the standard definition of velocity bias $b_v =
\sigma_{gal}/\sigma_{DM}$, where $\sigma_{gal}$ is the galaxy cluster
velocity dispersion, we can rewrite Equation~\ref{eqn_m1} as
\begin{equation}
b_v^{1/\alpha} M_{200} = 10^{15}\ M_{\sun} \frac{1}{h(z)}
\left(\frac{\sigma_{gal}}{\sigma_{15}}\right) ^{1/\alpha} \ , \label{eqn_m2}
\end{equation} 
where the quantity $b_v^{1/\alpha} M_{200}$ parameterizes our lack
of knowledge about velocity bias. Substituting in the measured values
for $\sigma_{gal}$ we obtain $b_v^{1/\alpha} M_{200} = 5.79^{+2.22}_{-1.99}
\times 10^{14} M_{\sun}$.

\citet[][and references therein]{bayliss11} discuss an ``orientation bias'' 
effect which causes an upward bias in the measured velocity dispersions of 
lensing-selected clusters, due to the higher likelihood of the alignment 
along the line of sight of the major axes of the cluster halos, which are in 
general triaxial.  \cite{bayliss11} estimate that on average this will result
in the dynamical mass estimate being biased high by 19-20\%, using the same 
relation between $M_{200}$ and velocity dispersion as we have used 
\citep[Eqn.~\ref{eqn_m1} above;][]{evrard08}.  Correcting for this 
orientation bias effect would result in 
$b_v^{1/\alpha} M_{200} = 4.8 \times 10^{14} M_{\sun}$, which is not a 
significant difference, as the change is well under $1\sigma$.  We therefore 
do not apply this correction, but we do note that it would improve the 
already good agreement with our other mass estimates in 
\S\ref{sec:cluster_properties} and \S\ref{sec:combined} (assuming no velocity
bias, $b_v = 1$.)

\section{Strong Lensing Properties}\label{sec:lens_modeling}

\begin{figure}
\plotone{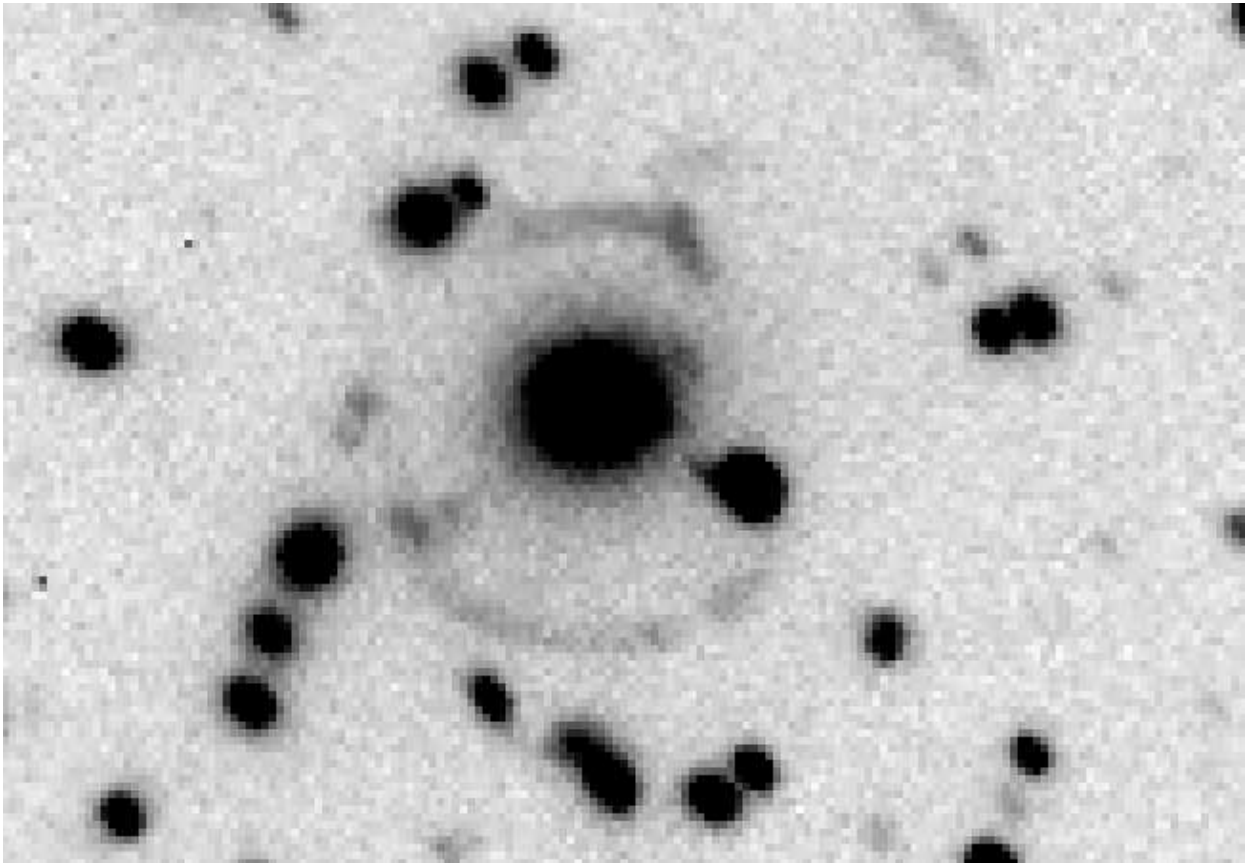}
\caption{The coadded $r$-band image. The lensing features can be
clearly seen.\label{arc-r-band}}
\end{figure}

We use the coadded $r$-band image shown in Fig.~\ref{arc-r-band} to study the
strong lensing features of the system as it has the best
seeing and hence shows the most detail. To remove the contribution to the arc fluxes from nearby objects we used GALFIT \citep{peng02} to model the profiles of these objects (galaxies and stars) and then subtracted the model from the image. This was done for all four bands $griz$. These subtracted images are used for all determinations of arc fluxes and positions. A number of individual knots can be observed in the system along with the more elongated features. For example it appears that knot A1 is actually composed of two individual bright regions which are resolved by the Sextractor object extraction described below. Knot A2 also appears to have two components although these are not resolved by the object extraction so we treat them as one in the modeling. Even though the cluster is fairly massive we do not see
evidence for additional arc-like features outside of the central
circular feature. In this case we expect the mass of the lens to be well constrained by the image positions.

We use the criteria that to obtain multiple images the average surface mass density within the tangential critical curve must equal the critical surface mass density $\Sigma_{crit}$. The tangentially oriented arcs occur at approximately the tangential critical curves and so the radius of the circle $\theta_{arc}$ traced by the arcs provides a measurement of the Einstein radius $\theta_{EIN}$ \citep{bartelmann96}. The mass
$M_{EIN}$ enclosed
with the Einstein radius is therfore given by
\begin{equation}
M_{EIN} = \Sigma_{crit}\pi (D_l \theta_{EIN})^{2} \label{eqn_M}
\end{equation}
Substituting for $\Sigma_{crit}$ gives
\begin{equation}
M_{EIN} = \frac{c^2}{4G} \frac{D_l D_s}{D_{ls}}\theta^2_{EIN} \label{eqn_SIS}
\end{equation}
where $D_{s}$ is the angular diameter
distance to the source, $D_{l}$ the angular diameter distance to the lens, and
$D_{ls}$ the angular diameter distance between the lens and the
source. These values are  $D_{s} = 1610$ Mpc, $D_{l} = 1081$ Mpc and
$D_{ls} = 825$ Mpc. 

To determine the Einstein radius we ran Sextractor \citep{bertin96} on the  $r$-band image. This identified eight distinct objects in the image. We used the coordinates of those eight objects and fit them to a circle. The radius of the circle gives us a measure of the Einstein radius. The Einstein radius we measure is $\theta_{EIN}=7.53\pm 0.25\arcsec$
which translates to $39.5\pm 1.3$ kpc. This yields
a mass estimate of $(1.5\pm 0.1) \times 10^{13}M_{\sun}$ and a
corresponding velocity dispersion (assuming an isothermal model for the mass distribution) of $\sigma = 694\pm 12 \ {\rm km \
s^{-1}}$.

The magnification of the lens $f_{lens}$ can be roughly estimated under the assumption that the 1/2-light radius of a source
at redshift $z\sim 0.9$ is about $0.46\arcsec$ (derived from the mock galaxy catalog described in \cite{jouvel09}). The ratio of the area subtended by the ring to that subtended by the source is $\sim 0.6 \times (4R/\delta r)$, where $R$ is the ring radius and $\delta r$ is the 1/2-light radius of the source. The $0.6$ factor accounts for the fraction of the ring that actually contains images. This gives a magnification of $f_{lens} = 39$. 

To obtain a more quantitative value for the magnification we have used the {\tt PixeLens}\footnote{Version 2.17: http://www.qgd.uzh.ch/programs/pixelens/} program \citep{saha04} to model the lens. {\tt PixeLens} is a parametric modeling program that reconstructs a pixelated mass map of the lens. It uses as input the coordinates of the extracted image positions and their parities along with the lens and source redshifts. It samples the solution space using a Markov Chain Monte Carlo method and generates an ensemble of mass models that reproduce the image positions. We used the Sextractor image positions obtained above and assigned the parities according to the prescription given in \cite{read07}. In \cite{saha04} they note that if one uses pixels that are too large then the mass distribution is poorly resolved and not enough steep mass models are allowed. We have chosen a pixel size such that this should not be a problem.

It is well known (see for example \cite{saha06}) that changing the slope of the mass profile changes the overall magnification, in particular a steeper slope produces a smaller magnification but does not change the image positions. Therefore the quoted magnification should be taken as a representative example rather than a definitive answer.
The magnification quoted is the sum over the average values of the magnification for each image position for 100 models. We obtain a value of $f_{lens} = 141\pm 39$ where the error is the quadrature sum of the RMS spreads of the individual image magnifications. {\tt PixeLens} can also determine the enclosed mass within a given radius. For the 100 models we obtain $M_{EIN} = (1.4\pm 0.02) \times 10^{13} M_{\sun}$ which is within $1\sigma$ of the mass obtained from the circle fit described above.

In order to combine the strong lensing mass with the mass estimate
from the weak lensing analysis (in \S\ref{sec:combined} below) we will need to estimate the mass within
$\theta_{EIN}$ that is due to dark matter alone ($M_{DM}$). To do this
we will need to subtract estimates of the stellar mass ($M_S$) and the hot gas mass
($M_G$) from the total mass $M_{EIN}$. 
To determine $M_S$ we use the GALAXEV \citep{bc03} evolutionary stellar 
population synthesis code to fit galaxy spectral energy distribution 
models to the $griz$ magnitudes of the BCG within the Einstein radius.
The BCG photometric data are taken from the GALFIT modeling described
above, and we sum up the light of the PSF-deconvolved
GALFIT model inside the Einstein radius.  The GALAXEV models considered are 
simple stellar population (SSP) models which have an initial, instantaneous 
burst of star formation;  such models provide good fits to early-type 
galaxies, such as those in clusters.  In particular we find a good fit
to the BCG, using a SSP model with solar metallicity, a \cite{chabrier03}
stellar initial mass function (IMF), and an age 9.25~Gyr (this age provided
the best $\chi^2$ over the range we considered, from 1~Gyr to 9.3~Gyr, 
the latter being the age of the universe for our cosmology at the cluster 
redshift $z = 0.38$).
The resulting stellar mass (more precisely the total stellar mass integrated
over the IMF) is $M_S = 1.7 \times 10^{12} M_{\sun}$.

To estimate the gas mass $M_G$ we have looked at estimates of hot gas fraction
$f_{gas}$ in cluster cores from X-ray observations. Typical $f_{gas}$ measurements
are of order 10\% \citep{maughan04,pointecouteau04} which give
us an $M_G$ estimate of $1.5\times 10^{12} M_{\sun}$.

Finally we calculate the {\em total} $M/L$ ratio within $\theta_{EIN}$ for
the $i$-band. This yields a value of $(M/L)_i = 33.7\pm 4.4
\ (M/L)_{\sun}$.

\section{Weak Lensing Measurements}\label{sec:weak_lens}

\subsection{Adaptive Moments}\label{sec:adaptive_moments}

We used the program Ellipto \citep{smith01} to compute adaptive moments
\citep{bernstein02,hirata04} of an object's light distribution, i.e., moments 
optimized for signal-to-noise via weighting by an elliptical Gaussian function 
self-consistently matched to the object's size.  Ellipto computes 
adaptive moments using an iterative method and runs off of an existing 
object catalog produced by SExtractor for the given image.  Ellipto is also a 
forerunner of the adaptive moments measurement code used in the 
SDSS photometric processing pipeline Photo.

We ran Ellipto on our coadded BCS images and corresponding SExtractor 
catalogs, doing so independently in each of the $griz$ filters to obtain
four separate catalogs of adaptive second moments:
\begin{eqnarray}
Q_{xx} & = & \int x^2 \ w(x,y) I(x,y) \ dx dy 
             \left/ \int w(x,y) I(x,y) \ dx dy \right. \\
Q_{yy} & = & \int y^2 \ w(x,y) I(x,y) \ dx dy
             \left/ \int w(x,y) I(x,y) \ dx dy \right. \\
Q_{xy} & = & \int x y \ w(x,y) I(x,y) \ dx dy 
             \left/ \int w(x,y) I(x,y) \ dx dy \right. \ ,
\end{eqnarray}
where $I(x,y)$ denotes the measured counts of an object at 
position $x,y$ on the CCD image, and $w(x,y)$ is the elliptical Gaussian
weighting function determined by Ellipto.  The images are oriented with the 
usual convention that North is up and East is to the left, i.e., 
right ascension increases along the $-x$ direction and declination 
increases along the $+y$ direction.  We then computed the ellipticity 
components $e_1$ and $e_2$ of each object using one of the standard definitions
\begin{eqnarray}
e_1 & = & (Q_{xx} - Q_{yy}) / (Q_{xx} + Q_{yy}) \\
e_2 & = & 2 Q_{xy} / (Q_{xx} + Q_{yy}) \ .
\end{eqnarray}

\subsection{PSF Modeling}\label{sec:psf_modeling}

\begin{figure}
\plotone{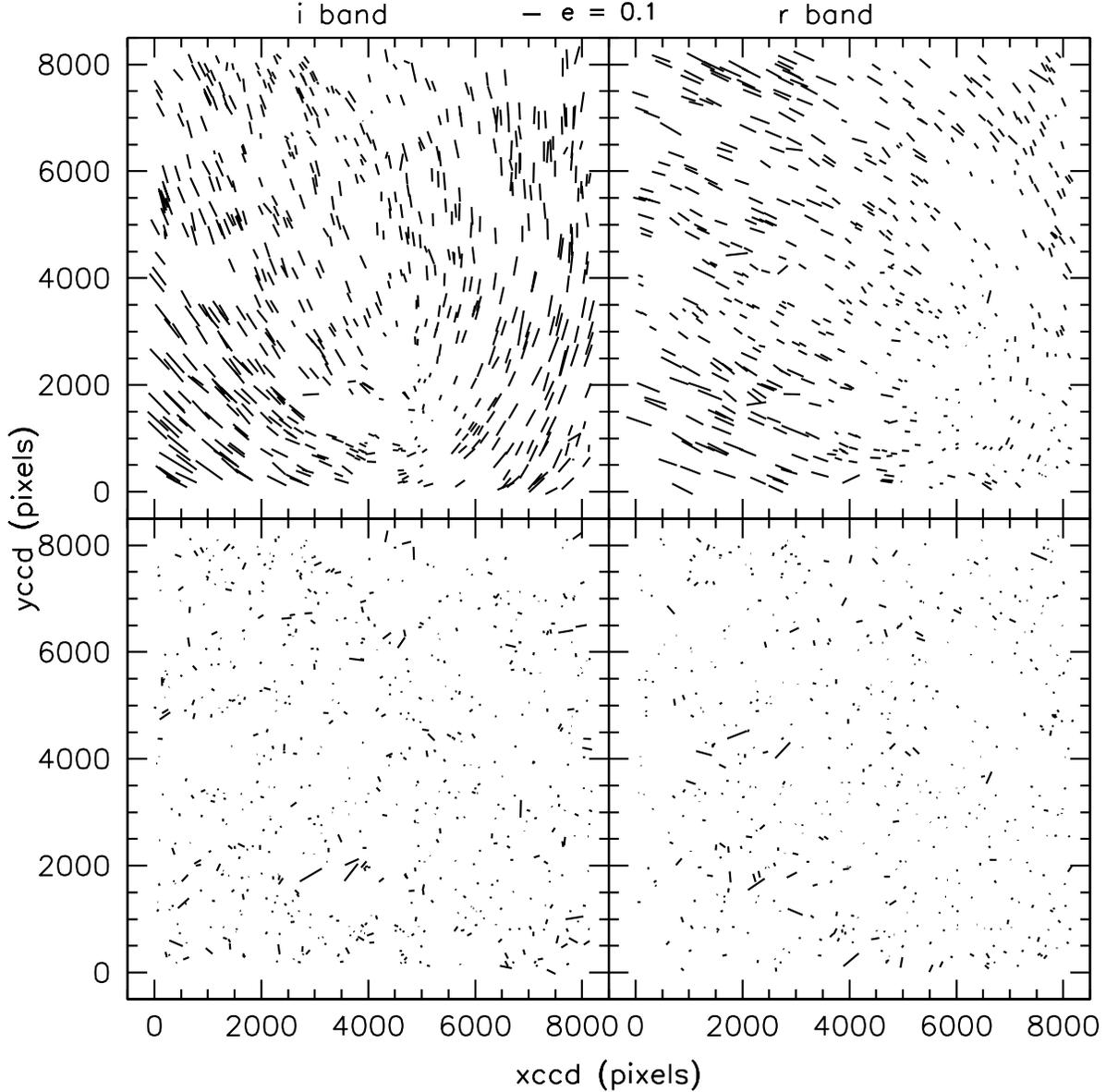}
\caption{({\em Top panels}) ``Whisker'' plots that show the clear spatial
variation of the PSF ellipticity vs.\ CCD $x,y$ position in our $i$- 
({\em left}) and $r$-band ({\em right}) images.  
The size of each whisker is proportional
to the PSF ellipticity $e_{PSF} = \sqrt{e_{1,PSF}^2 + e_{2,PSF}^2}$, where
a whisker with ellipticity $e = 0.1$ is shown at the top center of the figure.
Each whisker is oriented at an angle 
$\theta_{PSF} = \frac{1}{2} \tan^{-1}(e_{2,PSF}/e_{1,PSF})$
counterclockwise from horizontal.
({\em Bottom panels}) The corresponding whisker plots after subtraction of the 
PSF model described in \S\ref{sec:psf_modeling}, showing the removal of
the bulk of the spatial variation of the PSF ellipticities.
\label{fig_psf_whiskers}}
\end{figure}

For each filter, we then identified a set of bright but unsaturated stars
to use for PSF fitting.  We chose the stars from the stellar 
locus on a plot of the size measure $Q_{xx}+Q_{yy}$ from Ellipto 
vs.\ the magnitude {\tt MAG\_AUTO} from SExtractor, using simple cuts on
size and magnitude to define the set of PSF stars.  We then derived
fits of the ellipticities $e_1, e_2$ and the size $Q_{xx}+Q_{yy}$ of the 
stars vs.\ CCD $x$ and $y$ position, using polynomial functions 
of cubic order in $x$ and $y$ (i.e., the highest order terms 
are $x^3, x^2 y, xy^2$, and $y^3$).  On each image, these fits were done 
separately in each of 8 rectangular regions, defined by splitting the 
image area into 2 parts along the $x$ direction and into 4 parts
along the $y$ direction, corresponding to the distribution of the 8 Mosaic-II
CCDs over the image.  This partitioning procedure was needed in order
to account for discontinuities in the PSF ellipticity and/or size 
as we cross CCD boundaries in the Mosaic-II camera.  Also note that the
individual exposures comprising the final coadded image in each filter 
were only slightly dithered, so that the CCD boundaries were basically 
preserved in the coadd.  To illustrate the PSF variation in our images,
we present in Figure~\ref{fig_psf_whiskers} 
``whisker plots'' that show the spatial variation of the magnitude and 
orientation of the PSF ellipticity across our $i$- and $r$-band images .  
In addition, we also show the residuals in the PSF whiskers remaining after
our fitting procedure, showing that the fits have done a good job of 
modeling the spatial variations of the PSF in our data.

We next used our PSF model to correct our galaxy sizes and ellipticities
for the effects of PSF convolution.  Specifically, for the size measure 
$Q_{xx}+Q_{yy}$ we used the simple relation \citep[cf.][]{hirata03}
\begin{equation}
Q_{xx,true}+Q_{yy,true} = (Q_{xx,observed}+Q_{yy,observed}) 
                        - (Q_{xx,PSF}+Q_{yy,PSF})
\end{equation}
to estimate the true size $Q_{xx,true}+Q_{yy,true}$ of a galaxy from its 
observed size $Q_{xx,observed}+Q_{yy,observed}$, where $Q_{xx,PSF}+Q_{yy,PSF}$
is obtained from the PSF model evaluated at the $x,y$ position of the galaxy. 
For the ellipticities we similarly used the related expressions
\begin{eqnarray}
e_{i,true} & = & \frac{e_{i,observed}}{R_2} 
               + \left(1 - \frac{1}{R_2} \right) e_{i,PSF} \ , \ i=1,2 \label{eqn_etrue} \\
R_2 & \equiv & 1 - \frac{Q_{xx,PSF}+Q_{yy,PSF}}
                        {Q_{xx,observed}+Q_{yy,observed}}
\end{eqnarray}
The relations used in this simple correction procedure strictly hold only for 
unweighted second moments, or for adaptive moments in the special case when 
both the galaxy and the PSF are Gaussians.  We have therefore also checked the
results using the more sophisticated ``linear PSF correction'' procedure of 
\cite{hirata03}, which uses additional fourth order adaptive moment 
measurements (also provided here by Ellipto) in the PSF correction procedure.
In particular, the linear PSF correction method is typically applied in weak 
lensing analyses of SDSS data.  However, we found nearly 
indistinguishable tangential shear profiles from applying the two PSF 
correction methods, and we therefore adopted the simpler correction
method for our final results.

\subsection{Shear Profiles and Mass Measurements}\label{sec:shear_profiles} 

Given the estimates of the true galaxy ellipticities from 
Equation~(\ref{eqn_etrue}), we then computed the tangential ($e_T$) and 
B-mode or cross ($e_\times$) ellipticity components, in
a local reference frame defined for each galaxy relative to the BCG:
\begin{eqnarray}
     e_T & = & e_1 \cos(2\phi) - e_2 \sin(2\phi) \\
e_\times & = & e_1 \sin(2\phi) + e_2 \cos(2\phi)
\end{eqnarray}
where $\phi$ is the position angle (defined West of North) 
of a vector connecting the BCG to the galaxy in question.
Here we have dropped the subscript {\em true} for brevity.
The ellipticities were then converted to shears $\gamma$ 
using $\gamma = e / R$,  where $R$ is the responsivity, for which
we adopted the value $R = 2(1-\sigma^2_{SN}) = 1.73$, using
$\sigma_{SN} = 0.37$ as the intrinsic galaxy shape noise
as done in previous SDSS cluster weak lensing analyses
\citep[e.g.,][]{kubo07,kubo09}.

We then fit our galaxy shear measurements to an NFW profile by minimizing
the following expression for $\chi^2$:
\begin{equation}
\chi^2 = \sum_{i=1}^{N} \frac{[\gamma_i - \gamma_{NFW}(r_i; M_{200},c_{200})]^2}
                             {\sigma_\gamma^2} \label{eqn_chi2}
\end{equation}
where the index $i$ refers to each of the $N$ galaxies in a given
sample, $r_i$ is a galaxy's projected physical radius from
the BCG (at the redshift of the cluster), $\sigma_\gamma$ is the measured
standard deviation of the galaxy shears, and $\gamma_{NFW}$ is the 
shear given by Equations~(14-16) of \cite{wright00} for an NFW profile 
with mass $M_{200}$ and concentration $c_{200}$.  We used a standard 
Levenberg-Marquardt nonlinear least-squares routine to minimize $\chi^2$
and obtain best-fitting values and errors for the parameters 
$M_{200}$ and $c_{200}$ of the NFW profile.  
Similar fits of the weak lensing radial shear profile to a parameterized
NFW model have often been used to constrain the mass distributions of galaxy
clusters \citep[e.g.,][]{king01,clowe01,kubo09,oguri09,okabe10}.
Note that we chose the above 
expression for $\chi^2$ since it does {\em not} require us to do
any binning in radius, but for presentation purposes below we will have
to show binned radial shear profiles compared to the NFW shear profiles
obtained from our binning-independent fitting method.

For the shear fitting analysis, we defined galaxy samples separately
in each of the four $griz$ filters using cuts on the magnitude {\tt MAG\_AUTO} 
and on the size $Q_{xx,observed}+Q_{yy,observed}$, as detailed in 
Table~\ref{table_weak_lensing}.  The bright magnitude cut was chosen to
exclude brighter galaxies which would tend to lie in the foreground
of the cluster and hence not be lensed, while the faint magnitude cuts were
set to the photometric completeness limit in each filter, as defined
by the turnover magnitude in the histogram of SExtractor {\tt MAG\_AUTO} values.
For the size cut, we set it so that only galaxies larger than about 1.5 times
the PSF size would be used, as has been
typically done in SDSS cluster weak lensing analyses 
\citep[e.g.,][]{kubo07,kubo09}.  Note that in order to properly normalize
the NFW shear profile to the measurements, we also need to calculate the
critical surface mass density $\Sigma_{crit}$, which depends on the redshifts
of the lensed source galaxies as well as the redshift of the lensing cluster; 
see Equations~(9,14) of \cite{wright00}.  To do this, we did not use any 
individual redshift estimates for the source galaxies in our analysis, but 
instead we calculated an effective value of $1/\Sigma_{crit}$ via an integral 
over the source galaxy redshift distribution published for
the Canada-France-Hawaii Telescope Legacy Survey \citep[CFHTLS;][]{ilbert06},
as appropriate to the magnitude cuts we applied in each of the $griz$ filters.

\begin{figure}
\plotone{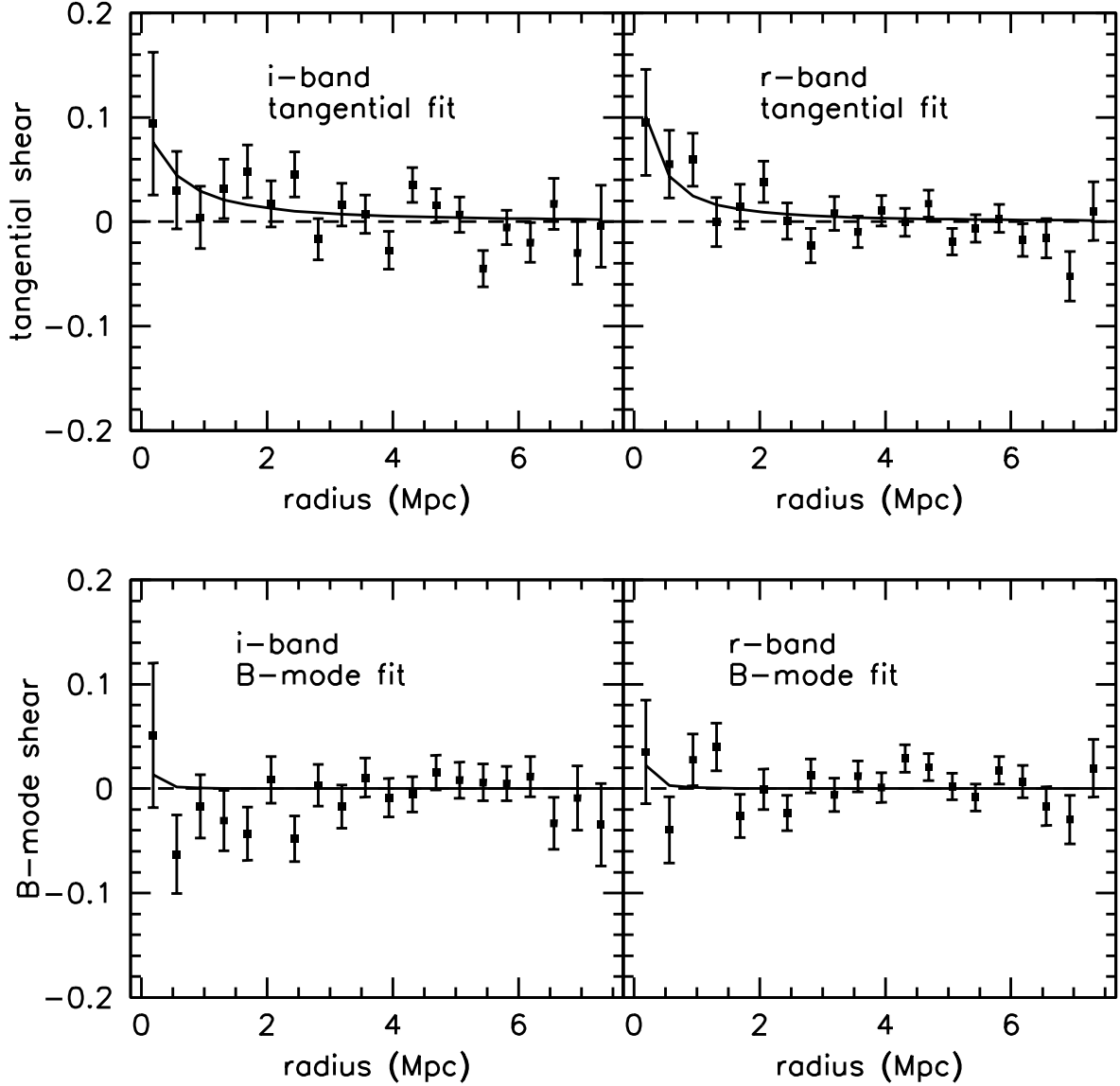}
\caption{The points with error bars show the tangential (top) 
and B-mode (bottom) radial shear profiles for the galaxy
sample used for weak lensing analysis in the $i$ (left) and $r$ (right) 
filters.
In each panel, the solid curve shows the shear profile for the best-fitting 
NFW mass density profile, as determined via the procedure described in 
\S\ref{sec:shear_profiles}.  The dashed horizontal lines indicate zero shear.
The best-fit NFW parameters and details of the galaxy sample are given 
in Table~\ref{table_weak_lensing}.
\label{nfw_fit_ir}}
\end{figure}

\begin{figure}
\plotone{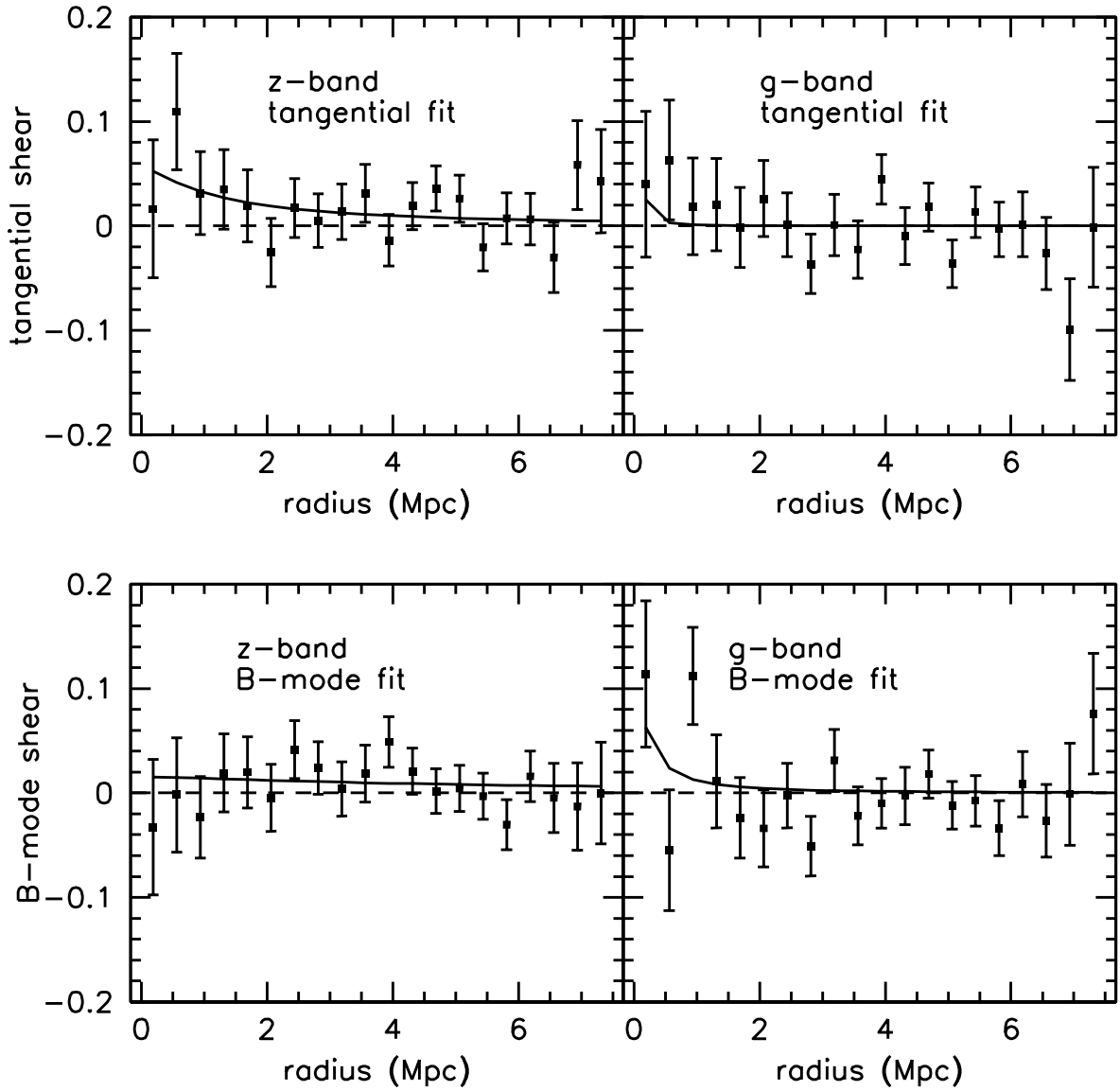}
\caption{Similar to Figure~\ref{nfw_fit_ir}, but for the $z$ (left)
and $g$ (right) filters.
\label{nfw_fit_zg}}
\end{figure}

Our NFW fitting results are shown in Figures~\ref{nfw_fit_ir}-\ref{nfw_fit_zg}
and detailed in Table~\ref{table_weak_lensing}.  We show results for both
the tangential and B-mode shear components.
As lensing does not produce a B-mode shear signal, these
results provide a check on systematic errors and should be consistent 
with zero in the absence of significant systematics.  For all of our filters,
our B-mode shear results are indeed consistent with no detected mass,
as the best-fit $M_{200}$ is within about 1$\sigma$ of zero.
On the other hand, for the tangential shear results in the $r$, $i$, and $z$ 
filters, we do indeed obtain detections of non-zero $M_{200}$ at the better 
than $1.5\sigma$ level.
In the $g$ filter we do not detect a non-zero $M_{200}$.
Comparing the weak lensing results from the different filters serves as a
useful check of the robustness of our lensing-based cluster mass measurement, 
in particular as the images in the different filters are subject to quite 
different PSF patterns, as shown earlier in Fig.~\ref{fig_psf_whiskers}.
Though the mass errors are large, the $M_{200}$ values from the $r$-, $i$-, 
and $z$-band weak lensing NFW fits are nonetheless consistent
with each other and with the masses derived earlier from the velocity
dispersion and maxBCG richness analyses.  Moreover, independent of the NFW fits, we have
also derived probabilities (of exceeding the observed $\chi^2$) that 
our {\em binned} shear profiles are consistent with the null hypothesis
of zero shear.  As shown in Table~\ref{table_weak_lensing}, we see that
the B-mode profiles are in all cases consistent with 
zero, as expected, but that the tangential profiles for the $r$ and $i$ 
filters are not consistent with the null hypothesis at about the $2\sigma$ 
level (probabilities $\approx 0.06$), thus providing model-independent 
evidence for a weak lensing detection of the cluster mass.

\subsection{Combining Weak Lensing Constraints from Different Filters}\label{sec:combined_filters} 

Here we will combine the weak lensing shear profile information from
the different filters $griz$ in order to improve the constraints on 
the NFW parameters, in particular on $M_{200}$.
The main complication here is that although the ellipticity measurement
errors are independent among the different filters, the most important
error for the shear measurement is the intrinsic galaxy shape noise,
which is correlated among filters because a subset of the galaxies
is common to two or more filters, and for these galaxies we expect their 
shapes to be fairly similar in the different filters. 
In particular we find that the covariance of the true galaxy ellipticities
between filters is large, for example, the covariance of $e_1$ between
the $i$ and $r$ filters, 
${\rm Cov}(e_{1,i},e_{1,r}) = \frac{1}{N} \sum (e_{1,i}-\bar{e}_{1,i})
(e_{1,r}-\bar{e}_{1,r})$, is about 0.9 times the variance of $e_1$ in the 
$i$ and $r$ filters individually.  The same holds true for $e_2$ and for the 
other filters as well.  We will not attempt to use a full covariance 
matrix approach to deal with the galaxy shape correlations when we combine 
the data from two or more filters.  Instead, we take a simpler approach of 
scaling the measured standard deviation of the
shear (the $\sigma_\gamma$ used to calculate $\chi^2$ in 
Equation~\ref{eqn_chi2}) by $\sqrt{N/N_{unique}}$, where $N$ is
the total number of galaxies in a given multi-filter sample,
and $N_{unique}$ is the number of unique galaxies in the same sample.
This is equivalent to rescaling $\chi^2$ in the NFW fit to correspond
to $N_{unique}$ degrees of freedom instead of $N$.
We have verified using least-squares fits to Monte Carlo simulations of 
NFW shear profiles that this simple approach gives the correct fit
uncertainties on $M_{200}$ and $c_{200}$ when the mock galaxy data contain
duplicate galaxies, with identical $e_1$ and $e_2$ values, simulating
the case of {\em completely} correlated intrinsic galaxy shapes among filters.
Note that our approach is conservative and will slightly 
overestimate the errors, because the galaxy shapes in the real data
are about 90\% correlated, not fully correlated, among filters.

\begin{figure}
\plotone{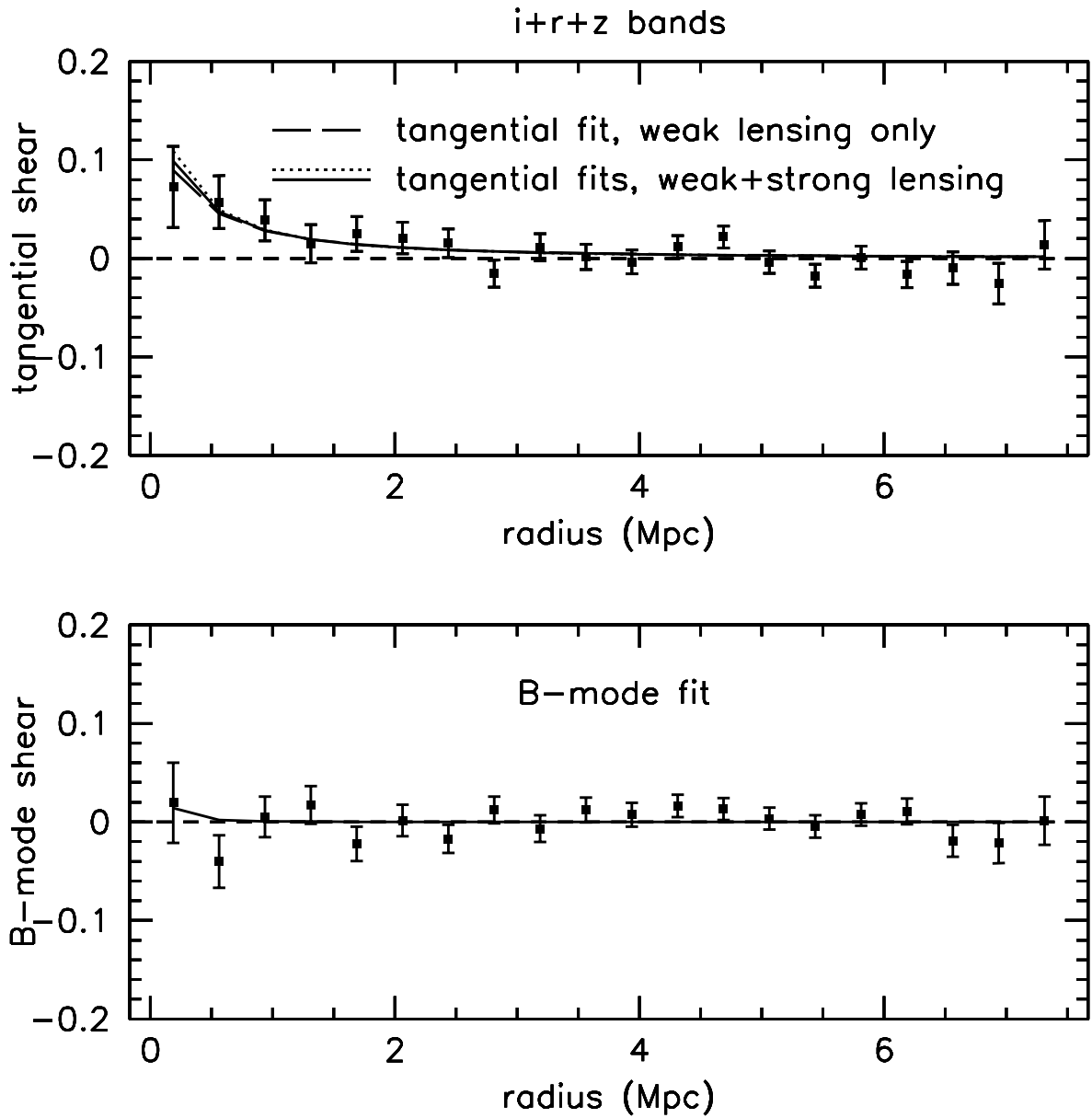}
\caption{Similar to Figure~\ref{nfw_fit_ir}, but for the multi-filter $i+r+z$
         sample.  For the tangential shear profile fits in the top panel,
         the long-dashed curve gives the results using weak lensing only, while
         the dotted and solid curves give the results using combined weak plus
         strong lensing.  The dotted curve is for the case where we estimated
         the dark matter mass within the Einstein radius by subtracting off 
         just a stellar mass contribution, while the solid curve is for the
         case where we also subtracted off an estimated gas mass contribution.
         See \S\ref{sec:combined_filters}, \S\ref{sec:combined},
         and Table~\ref{table_weak_lensing} for details.
\label{nfw_fit_irz_sl}}
\end{figure}

Before fitting the combined shear data from multiple filters, we
make one additional multiplicative rescaling of the shear values, 
so that all filters will have the same effective value of $1/\Sigma_{crit}$,
corresponding to a fiducial effective source redshift $z_{crit} = 0.7$.
This correction is small, with the largest being a factor of 1.18 for
the $z$-band data.  The results of the NFW fits for the multi-filter samples
are given in Table~\ref{table_weak_lensing}, where we have tried the
filter combinations $i+r$, $i+r+z$, and $i+r+z+g$.  We see that these
multi-filter samples all provide better fractional errors on $M_{200}$ 
compared to those from the single-filter data.  Also, as expected,
the B-mode results in all cases are consistent with no detected 
$M_{200}$ and zero shear.  For our final weak lensing
results, we adopt the NFW parameters from the $i+r+z$ sample, as it
provides the best fractional error ($\sigma_{M_{200}}/M_{200} \approx 0.5$) 
on $M_{200}$;  we obtain $M_{200} = 5.0^{+2.9}_{-2.3} \times 10^{14}  M_\Sun$, and $c_{200} = 4.9^{+3.9}_{-2.2}$.  
Figure~\ref{nfw_fit_irz_sl} shows the shear 
profile data and best fit results for the $i+r+z$ sample.
This final weak lensing value for $M_{200}$ agrees
well with the earlier values of $M_{200}$ derived from the cluster
galaxy velocity dispersion (assuming no velocity bias) and from the cluster
richness $N_{200}$.

\section{Combined Constraints on Cluster Mass and Concentration}\label{sec:combined_constraints}

\subsection{Combining Strong and Weak Lensing}\label{sec:combined}

In this section we combine the strong lensing and weak lensing
information together in order to further improve our constraints
on the NFW profile parameters, in particular on the concentration 
parameter $c_{200}$.  The addition of the strong lensing information provides 
constraints on the mass within the Einstein radius, close to the cluster 
center, thereby allowing us to better measure the central concentration
of the NFW profile and improve the uncertainties on the concentration $c_{200}$.
\cite{oguri09} incorporated the strong lensing information in the form
of a constraint on the Einstein radius due to just the dark matter 
distribution of the cluster, and they specifically excluded the contribution
of (stellar) baryons to the Einstein radius.  Their intent, as well as ours
in this paper (\S~\ref{sec:all_combined}), is to compare the observed
cluster NFW concentration to that predicted from dark-matter-{\em only} 
simulations.  Thus the contribution of baryonic matter should be removed,
most importantly in the central region within the Einstein radius, where
baryonic effects are the largest due in particular to the presence of
the BCG.  In practice with the present data we can do this separation
of the baryonic contribution only for the strong lensing constraint, and 
strictly speaking the weak lensing profile results from the total mass 
distribution rather than from dark matter alone.

Here we combine the strong and weak lensing data using an analogous
but somewhat simpler method compared to that of \cite{oguri09}, specifically by
adding a second term to $\chi^2$ (Equation~\ref{eqn_chi2}) that describes the 
constraint on the dark matter (only) mass within the observed Einstein 
radius:
\begin{equation}
\chi^2 = \sum_{i=1}^{N} \frac{[\gamma_i - \gamma_{NFW}(r_i; M_{200},c_{200})]^2}
                             {\sigma_\gamma^2}
       + \frac{[M_{DM}(< \theta_E) - M_{NFW}(< \theta_E; M_{200},c_{200})]^2}
              {\sigma_{M_{DM}(< \theta_E)}^2} \label{eqn_chi2_wl_sl}
\end{equation}
where $\theta_E = 7.53\arcsec$ is the observed Einstein radius due to the 
{\em total} cluster mass distribution, $M_{DM}(< \theta_E)$ is the 
dark matter (only) mass within $\theta_E$, and 
$M_{NFW}(< \theta_E; M_{200},c_{200})$ is the mass within $\theta_E$ of an 
NFW profile with mass $M_{200}$, concentration $c_{200}$, redshift $z = 0.38$, 
and source redshift $z = 0.9057$.  
$M_{NFW}(< \theta_E; M_{200},c_{200})$ is derived based on Equation~(13) of 
\cite{wright00}.  
As obtained earlier in \S\ref{sec:lens_modeling}, we 
estimate $M_{DM}(< \theta_E)$ by subtracting estimates of the stellar mass and
hot gas mass from the total mass within $\theta_E$, obtaining
$M_{DM}(< \theta_E) = (1.18 \pm 0.2) \times 10^{13} M_\Sun$ when subtracting 
off both stellar and gas mass, 
or $M_{DM}(< \theta_E) = (1.33 \pm 0.2) \times 10^{13} M_\Sun$ when 
subtracting off only stellar mass.  The former is our best estimate of 
$M_{DM}(< \theta_E)$, while the latter serves as an upper limit on
$M_{DM}(< \theta_E)$ and hence on the best-fit concentration $c_{200}$.
We also conservatively estimate the error on $M_{DM}(< \theta_E)$
to be one of the stellar mass/gas mass components added in quadrature to 
the uncertainty on the total $M_{EIN}$ from \S\ref{sec:lens_modeling}.

We apply the combined strong plus weak lensing analysis to our best
weak lensing sample, the multi-filter $i+r+z$ data set.  The fit
results are given in Table~\ref{table_weak_lensing} and shown in 
Figure~\ref{nfw_fit_irz_sl}.  We find $M_{200} = 4.9^{+2.9}_{-2.2} \times 10^{14}$
solar masses, nearly identical to the final weak lensing result.
We also get a concentration $c_{200} = 5.5^{+2.7}_{-1.6}$, again consistent 
with the final weak lensing fit, but with a 30\% improvement in
the error on $c_{200}$, demonstrating the usefulness of adding the strong lensing
information to constrain the NFW concentration.  
Using the upper limit $M_{DM}(< \theta_E)$ value (with only stellar mass 
subtracted) gives nearly the same $M_{200} = 4.8^{+2.8}_{-2.2} \times 10^{14} M_\Sun$,
while the resulting NFW concentration is higher, as expected, 
with $c_{200} = 6.2^{+3.2}_{-1.7}$, but still consistent with the fit using 
our best estimate of $M_{DM}(< \theta_E)$.

\subsection{Combining Lensing, Velocity Dispersion and Richness Constraints}\label{sec:all_combined}

In the above sections we have obtained quite consistent constraints on the 
cluster mass $M_{200}$ using three independent techniques:  
(1) $M_{200}({\rm lensing}) = 4.9^{+2.9}_{-2.2} \times 10^{14} M_\Sun$ from 
combined weak + strong lensing (\S\ref{sec:combined});  
(2) $M_{200}(\sigma_c) = 5.79^{+2.22}_{-1.99} \times 10^{14} M_\Sun$ from the 
cluster galaxy velocity dispersion $\sigma_c$ (\S\ref{sec:veldisp}; 
assuming no velocity bias, $b_v = 1$); and 
(3) $M_{200}(N_{200}) = (4.6 \pm 2.1) \times 10^{14} M_\Sun$ from the 
maxBCG-defined cluster richness $N_{200}$ (\S\ref{sec:cluster_properties}).
We note that these methods are subject to different assumptions and systematic
errors.  For example, the velocity dispersion based mass estimate assumes
the cluster is virialized, an assumption supported by the Gaussian-shaped 
velocity distribution of the cluster members shown in Fig.~\ref{vel_disp}.
Also, the richness based mass estimate relies on the $N_{200}$-$M_{200}$
calibration \citep{johnston07} obtained for SDSS maxBCG clusters at lower
redshifts $z = 0.1-0.3$ and assumes that this calibration remains valid for 
our cluster at $z = 0.38$.  It is encouraging that we are obtaining a cluster
mass measurement that appears to be robust to these disparate assumptions
and that shows good agreement among multiple independent methods.

\begin{figure}
\plotone{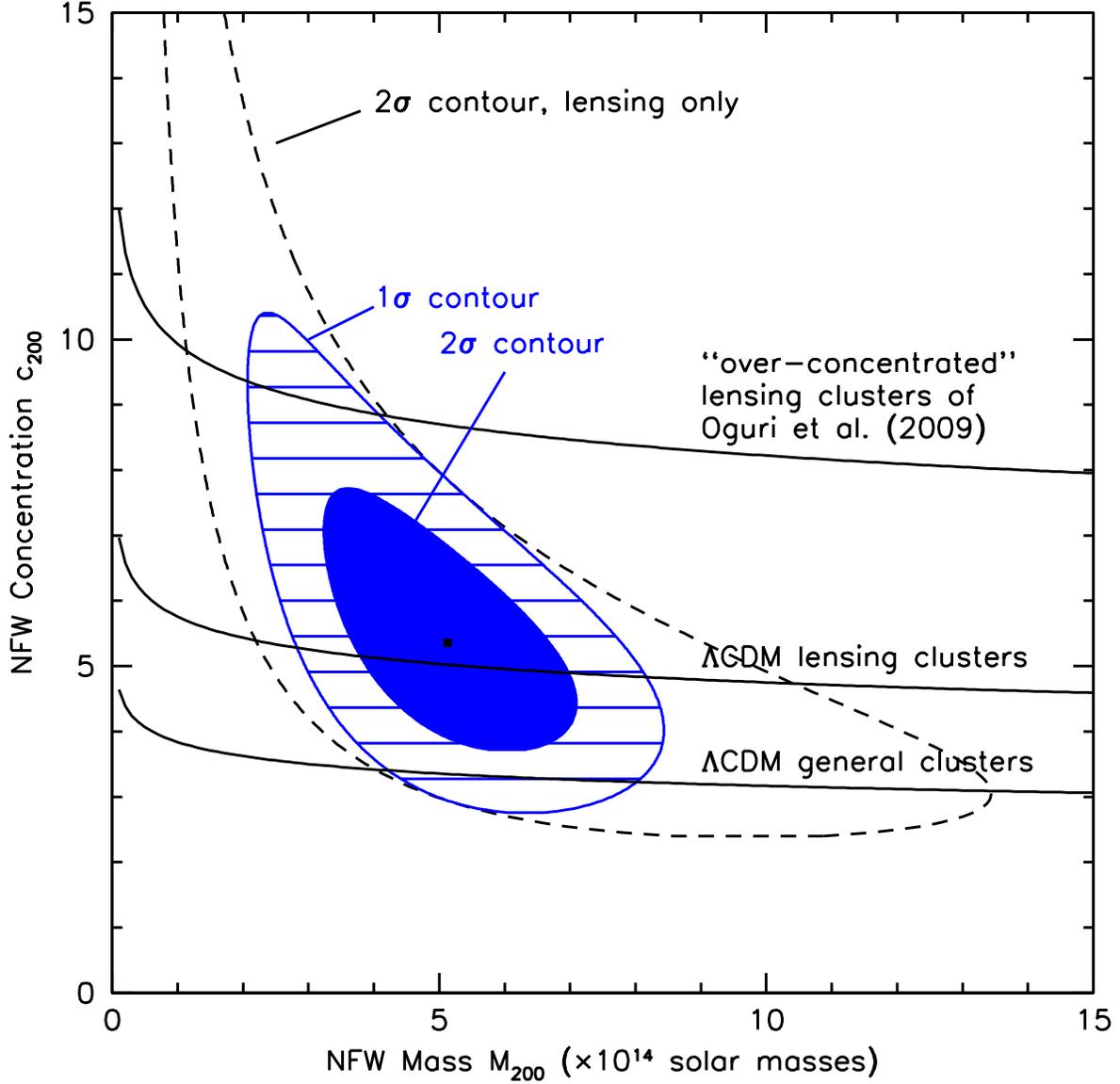}
\caption{Confidence contours for the best-fitting NFW mass $M_{200}$ and
concentration $c_{200}$, obtained by combining the lensing, velocity dispersion,
and cluster richness constraints, as described in \S\ref{sec:all_combined}.
The 2-parameter, $1\sigma$ contours are shown in solid blue, while
the $2\sigma$ contours are shown in hatched blue.  The outer dashed contours
show the 2-parameter, $2\sigma$ constraints derived solely from the 
weak + strong lensing analysis of \S\ref{sec:combined}.
Also, as described in \S\ref{sec:all_combined}, the 3 mostly horizontal curves 
show the concentration vs.\ mass relation at $z = 0.4$ for: 
(bottom) clusters overall from $\Lambda$CDM simulations; 
(middle) lensing selected clusters from simulations; 
and (top) a real lensing cluster sample from \cite{oguri09}.
\label{contours_M200_c}}
\end{figure}

We will therefore combine the results from the different techniques in
order to obtain final constraints on $M_{200}$ and concentration $c_{200}$
that are significantly improved over what any one technique permits.
Specifically, we can add the $M_{200}$ constraints from the velocity dispersion
and richness measurements as additional terms to the weak + strong lensing
$\chi^2$ (Equation~\ref{eqn_chi2_wl_sl}):
\begin{eqnarray}
\chi^2 & = & \sum_{i=1}^{N} \frac{[\gamma_i - \gamma_{NFW}(r_i; M_{200},c_{200})]^2}
                             {\sigma_\gamma^2}
         + \frac{[M_{DM}(< \theta_E) - M_{NFW}(< \theta_E; M_{200},c_{200})]^2}
                    {\sigma_{M_{DM}(< \theta_E)}^2} \nonumber \\
       & + & \frac{[M_{200}(\sigma_c)-M_{200}]^2}{\sigma_{M_{200}(\sigma_c)}^2}
           + \frac{[M_{200}(N_{200})-M_{200}]^2}{\sigma_{M_{200}(N_{200})}^2}
\label{eqn_chi2_all}
\end{eqnarray}
Minimizing this overall $\chi^2$ results in the final best-fitting 
NFW parameters $M_{200} = 5.1^{+1.3}_{-1.3} \times 10^{14} M_\Sun$ 
and $c_{200} = 5.4^{+1.4}_{-1.1}$.  These results are consistent with the final 
lensing-based values $M_{200}({\rm lensing}) = 4.9^{+2.9}_{-2.2} \times 10^{14} M_\Sun$
and $c_{200}({\rm lensing}) = 5.5^{+2.7}_{-1.6}$, but have errors nearly a factor of two 
smaller.  Note these quoted errors are 1-parameter, $1\sigma$ uncertainties;
we plot the joint 2-parameter, $1\sigma$ and $2\sigma$ contours in 
Fig.~\ref{contours_M200_c}.

We also note that for the three methods weak lensing, velocity dispersion, 
and cluster richness, the corresponding NFW parameters result from 
the {\em total} mass distribution, consisting of both dark matter and baryonic 
(stellar plus hot gas) components.  Dark matter is dominant over the bulk
of the cluster, while baryons can have a significant effect in the cluster core
\citep[e.g.,][]{oguri09}.  As described earlier (\S~\ref{sec:combined}),
we have thus subtracted off the baryonic contribution to the strong lensing 
constraint as the intent is to compare (see below) our cluster concentration 
value against those from dark-matter-only simulations.  Note that we have not 
isolated the dark matter contribution for the other three methods
and cannot easily do so.  
For weak lensing, the shear profile is sensitive to the total mass 
distribution, not just to dark matter.  For the velocity dispersion method, 
the galaxies act as test particles in the overall cluster potential, which is 
due, again, to both dark matter and baryons.  For the cluster richness method,
the \cite{johnston07} $N_{200}$-$M_{200}$ relation we use was derived from
stacked cluster weak lensing shear profile fits, including a BCG contribution
but otherwise no other baryonic components;  thus again the $M_{200}$ value
is essentially for the total mass distribution.  Nonetheless, the bulk
of the baryonic contribution is in the cluster core and is accounted for
via the strong lensing constraint, so we expect the comparison
below of our cluster concentration value to those of dark matter simulations 
to be a reasonable exercise.

Recent analyses \citep[e.g.,][]{oguri09,broadhurst_barkana08} of strong lensing
clusters have indicated that these clusters are more concentrated than would 
be expected from $\Lambda$CDM predictions, though others have argued that
no discrepancy exists if baryonic effects are accounted for \citep{richard10}.
In the former case, \cite{oguri09} found a concentration 
$c_{\rm vir} \approx 9$ for the 
10 strong lensing clusters in their analysis sample, compared to a value
of $c_{\rm vir} \approx 6$ expected for strong-lensing-selected clusters
or $c_{\rm vir} \approx 4$ for clusters overall
\citep[e.g.,][]{broadhurst_barkana08,oguri_blandford09}.
We illustrate these different concentration values in 
Fig.~\ref{contours_M200_c}.  We use Eqn.~(17) of \cite{oguri09},
$\bar{c}_{\rm vir}({\rm sim}) = \frac{7.85}{(1+z)^{0.71}}(M_{\rm vir}/2.78 \times 10^{12} M_\Sun)^{-0.081}$,
which comes from the $\Lambda$CDM N-body simulations of \cite{duffy08}, to 
show the typical concentration of clusters overall, and multiply by a factor 
of 1.5 \citep{oguri09} to show the higher concentration expected for lensing
selected clusters.  We also use Eqn.~(18) of \cite{oguri09},
$\bar{c}_{\rm vir}({\rm fit}) = \frac{12.4}{(1+z)^{0.71}}(M_{\rm vir}/10^{15} M_\Sun)^{-0.081}$, to show the fit results for their cluster sample.
In these relations, we set $z = 0.4$ to match the redshift of our cluster.
Moreover, we convert from the $M_{\rm vir}, c_{\rm vir}$ convention used
by \cite{oguri09} to our $M_{200}, c_{200}$ convention, using the detailed
relations found in Appendix~C of \cite{hu03} or in the 
Appendix of \cite{johnston07}. For the plotted $M_{200}$ range, it turns out 
that $c_{200} \approx 0.83\ c_{\rm vir}$.
From Fig.~\ref{contours_M200_c}, we see that our best-fit value of 
$c_{200} = 5.4^{+1.4}_{-1.1}$ is most consistent with the nominal $\Lambda$CDM 
concentration value for lensing-selected clusters, and does not suggest the 
need for a concentration excess in this particular case.  
It's likely that larger strong lensing cluster samples will be needed
to more robustly compare the distribution of concentration values with 
the predictions of $\Lambda$CDM models.

\section{Source Galaxy Star Formation Rate}\label{sec:SFR}

We can use the [OII]3727 line in the calibrated
spectra described in \S~\ref{sec:redshifts} to estimate the star formation rate (SFR).  As noted by \citet{kennicutt98} the luminosities of forbidden lines like [OII]3727 are not directly coupled to the ionizing luminosity and their excitation is also sensitive to abundance and the ionization state of the gas. However the excitation of [OII] is well behaved enough that it can be calibrated through H$\alpha$ as an SFR tracer.  This indirect calibration is very useful for studies of distant galaxies because [OII]3727 can be observed out to redshifts $z\approx 1.6$ and it has been measured in several large samples of faint galaxies (see references in \citet{kennicutt98}). If we know the [OII] luminosity then we can use
equation 3 from \citet{kennicutt98} to determine a star formation rate for the galaxy
\begin{equation}
{\rm SFR} (M_\sun\ {\rm yr}^{-1}) = (1.4 \pm 0.4) \times 10^{-41} (L{\rm [OII]})({\rm ergs\ s}^{-1})\label{eqn_SFR}
\end{equation}
where the uncertainty reflects the range between blue emission-line galaxies (lower limit) and more luminous spiral and irregular galaxies (upper limit).

As noted above, in order to extract the
SFR we need to determine the total source flux from the [OII] line. We
determine this using 
\begin{equation}
f(\nu)_{[OII]} = \frac{f(\nu)_L}{f(\nu)_S}\times 
f(\nu)_I \label{eqn_SFR_flux}
\end{equation}
where $f(\nu)_{[OII]}$ is the total flux emitted by the source in the [OII]
line, $f(\nu)_L$ is the flux measured in the [OII] line in each spectrum,
$f(\nu)_S$ is the flux in the knot spectrum contained within the {\it
i}-band filter band pass and $f(\nu)_I$ is the flux from the
source in the {\it i}-band.

Using the GALFIT-subtracted {\it i}-band image we determine $f(\nu)_I$ by summing the flux in an annulus of width 3\arcsec\ that encompasses the arcs. The flux $f(\nu)_L$ is measured by fitting a gaussian plus a continuum to the [OII] line in each spectrum and integrating the flux under the gaussian fit. The flux $f(\nu)_S$ is calculated as follows. For each spectrum we first fit the continuum level, we then add the fitted continuum plus the [OII] line flux and convolve it with the filter response curve for the SDSS {\it i}-band filter and integrate the convolved spectrum.

We have determined $f(\nu)_{[OII]}$ separately for each knot that was targeted
for spectra. The fluxes are listed in Table~\ref{table_sfr} for each knot.
We convert $f(\nu)_{[OII]}$ into an [OII] luminosity and then use
Equation~\ref{eqn_SFR} to determine a star formation rate for each knot. This rate is the raw rate which must be scaled by the lens magnification $f_{lens}$ to determine the true rate. We quote the SFR for the two values of $f_{lens}$ that were determined in \S\ref{sec:lens_modeling}.
We assume one magnitude
of extinction \citep{kennicutt98} and have corrected the measured
[OII] luminosity to account for this.  This yields the
star formation rates listed in Table~\ref{table_sfr} for the two values of $f_{lens}$. The rate for knot A3 is higher by a factor of 2 compared to the others because it has a small $f(\nu)_S$ compared to the other knots but the value of $f(\nu)_L$ is quite similar to the other knots. This can clearly be seen in Figure~\ref{arc_spectra}. We can combine the measurements for the four knots using a simple average to quote an overall SFR. This yields values of
${\rm SFR}(f_{lens}=49) = 4.6 \pm 0.7$ and ${\rm SFR}(f_{lens}=141) = 1.3\pm0.2$.

These rates are significantly smaller that those obtained for the 8
o'clock arc \citep{allam07} and the Clone \citep{lin09} which were
$229 M_\sun\ {\rm yr}^{-1}$ and $45 M_\sun\ {\rm yr}^{-1}$
respectively (after converting to our chosen cosmology). Both these
systems were at much higher redshift (2.72 and 2.0 respectively) so one
would potentially expect higher rates from these systems. They also had smaller values of $f_{lens}$. We can
compare our result to blue galaxies at similar redshift from the
DEEP2 survey \citep{cooper08}. Using Figure 18 of \citet{cooper08} we
obtain a median SFR of about $34 M_\sun\ {\rm yr}^{-1}$ for a redshift
$z=0.9$ galaxy which is also higher than our measurement.  Other measurements using the AEGIS field
\citep{noeske07} give a median SFR ranging from $10 M_\sun\ {\rm yr}^{-1}$
to $40 M_\sun\ {\rm yr}^{-1}$ depending weakly on the galaxy mass, which is
unknown in our case. Our
measurement can be compared to the far-right plot of Figure 1 in
\citet{noeske07} and we fall on the low side of the measured data. Note that these conclusions are
dependent on the magnification values used, for example smaller values such as those obtained for the Clone or the 8 o'clock arc would yield larger values for the SFR.

\section{Conclusions}\label{sec:conclusions}

We have  reported on  the discovery of a star-forming galaxy at  a
redshift of $z=0.9057$ that is being strongly lensed by a
massive galaxy cluster at a redshift of $z=0.3838$.

The Einstein radius determined from the lensing features is
$\theta_{EIN}=7.53\pm 0.25\arcsec$ and the enclosed mass is
$(1.5\pm 0.1) \times 10^{13}M_{\sun}$, with a
corresponding SIS velocity dispersion of $\sigma = 694 \pm 12 \ {\rm km \
s^{-1}}$.

Using GMOS spectroscopic redshifts measured for 30 cluster member galaxies,
we obtained a velocity dispersion $\sigma_c = 855^{+108}_{-96} \ {\rm km \
s^{-1}}$ for the lensing cluster.

We have derived estimates of $M_{200}$ from measurements of (1) weak lensing, 
(2) weak + strong lensing, (3) velocity dispersion $\sigma_c$,
and (4) cluster richness $N_{200} = 55$.  We obtained the following
results for $M_{200}$: 
(1) $M_{200}({\rm weak \ lensing}) = 5.0^{+2.9}_{-2.3} \times 10^{14} M_\Sun$, 
(2) $M_{200}({\rm lensing}) = 4.9^{+2.9}_{-2.2} \times 10^{14} M_\Sun$,  
(3) $M_{200}(\sigma_c) = 5.79^{+2.22}_{-1.99} \times 10^{14} M_\Sun$
(assuming no velocity bias, $b_v = 1$), and 
(4) $M_{200}(N_{200}) = (4.6 \pm 2.1) \times 10^{14} M_\Sun$. These
results are all very consistent with each other. The combination of
the results from methods 2, 3 and 4 give $M_{200} = 5.1^{+1.3}_{-1.3}
\times 10^{14} M_\Sun$, which is fully consistent with the individual
measurements but with an error that is smaller by a factor of nearly two. The
final NFW concentration from the combined fit is $c_{200} = 5.4^{+1.4}_{-1.1}$,
which is also consistent with the lensing-based value but again with a 
smaller error. 

We have compared our measurements of $M_{200}$ and $c_{200}$ with
predictions for (a) clusters from $\Lambda$CDM simulations, (b)
lensing selected clusters from simulations, and (c) a real sample of
cluster lenses from \citet{oguri09}. We find that we are most
compatible with the predictions from $\Lambda$CDM simulations for lensing
clusters, and we see no evidence that an increased concentration is needed 
for this one system.
We are studying this further using other lensing clusters we observed from the
SDSS \citep{diehl09}. These clusters will be the subject of a future paper. 

Finally, we have estimated the star forming rate (SFR) to be between 1.3 to 4.6
$M_\sun \ {\rm yr}^{-1} $, depending on magnification. These are small star-forming rates 
when compared to some of our 
previously reported systems, and are also small when compared with 
rates found for other galaxies at similar redshifts. However we caution that this conclusion is entirely dependent on the derived lens magnification.



\acknowledgments
We thank the anonymous reviewer for helpful comments which have improved the paper.

Based on observations obtained at the Gemini Observatory, which is operated by the 
Association of Universities for Research in Astronomy, Inc., under a cooperative agreement 
with the NSF on behalf of the Gemini partnership: the National Science Foundation (United 
States), the Science and Technology Facilities Council (United Kingdom), the 
National Research Council (Canada), CONICYT (Chile), the Australian Research Council (Australia), 
Minist\'{e}rio da Ci\^{e}ncia e Tecnologia (Brazil) 
and Ministerio de Ciencia, Tecnolog\'{i}a e Innovaci\'{o}n Productiva (Argentina). Gemini Proposal ID GS-2007B-Q-228.

The Blanco Cosmology Survey was performed on the Blanco 4m telescope
located at the Cerro Tololo Inter-American Observatory (National Optical Astronomy
Observatory) which is operated by the Association of Universities for
Research in Astronomy, under contract with the National Science
Foundation.

S. S. Allam acknowledges support from an HST Grant. Support of program
no. 11167 was provided by NASA through a grant from the Space
Telescope Science Institute, which is operated by the Association of
Universities for Research in Astronomy, Inc., under NASA contract
NAS5-26555.

AZ and JM acknowledge the support of the Excellence Cluster Universe in
Garching. Support for M.B. was provided by the W. M. Keck Foundation. 

Thanks to Jeff Kubo for providing a copy of the Ellipto adaptive moments code.
Thanks also go to Quarknet students Liana Nicklaus, Gina Castelvecchi, 
Braven Leung, Nick Gebbia, Alex Fitch, and their advisor Patrick Swanson,
who worked with HL during summer 2009 on a weak lensing mass analysis
of this cluster with different codes than used here.

Fermilab is operated by Fermi Research Alliance, LLC under Contract
No. DE-AC02-07CH11359 with the United States Department of Energy.

\clearpage



\clearpage


\clearpage

\clearpage

\clearpage

\clearpage

\clearpage

\clearpage

\clearpage

\clearpage

\clearpage

\clearpage

\clearpage

\begin{deluxetable}{ccccl}
\tablewidth{0pt}
\tablecaption{Observation Log\label{table_obslog}}
\tablehead{
\colhead{Filter/Grating} & 
\colhead{UT Date}     &
\colhead{Exposure}    & 
\colhead{Seeing}   &
\colhead{Notes}}
\startdata
\sidehead{BCS Imaging}
 $g$ & 14 Dec 2006 & 2$\times 125$ sec & $1.44\arcsec$ & \\
 $r$ & 14 Dec 2006 & 2$\times 300$ sec & $1.29\arcsec$ & \\
 $g$ & 11 Nov 2008 & 2$\times 125$ sec & $1.03\arcsec$ & \\
 $r$ & 11 Nov 2008 & 2$\times 300$ sec & $0.88\arcsec$ & \\
 $i$ & 30 Oct 2006 & 3$\times 450$ sec & $1.18\arcsec$ & \\
 $z$ & 30 Oct 2006 & 3$\times 450$ sec & $1.31\arcsec$ & \\
\sidehead{GMOS spectroscopy}
GG455 & 4 Aug 2007 & 4$\times 900$ sec & $0.56\arcsec$ & Mask 1
includes knots A1,A3,A4 \\
GG455 & 4 Aug 2007 & 4$\times 900$ sec & $1.14\arcsec$ & Mask 2
includes BCG and knot A2 \\
GG455 & 4 Aug 2007 & 1$\times 5$ sec & - & Cu-Ar Mask 1 \\
GG455 & 4 Aug 2007 & 1$\times 5$ sec & - & Cu-Ar Mask 2 \\
GG455 & 14 Aug 2007 & 1$\times 5$ sec & - & $1.5\arcsec$ slit \\
GG455 & 14 Aug 2007 & 1$\times 90$sec & $0.95\arcsec$ & Standard star
EG21 \\ 
\enddata
\end{deluxetable}

\clearpage

\begin{deluxetable}{cccccc}
\tabletypesize{\scriptsize}
\tablewidth{0pt}
\tablecaption{Knots Targeted for Spectroscopy\label{table_knots}}
\tablehead{
\colhead{Knot} &
\colhead{RA\tablenotemark{a}} &
\colhead{Dec\tablenotemark{a}} &
\colhead{$i(3\arcsec)$\tablenotemark{b}} &
\colhead{$g-r$\tablenotemark{c}} &
\colhead{$r-i$\tablenotemark{c}}}
\startdata
A1 & 357.912477 & -54.881691 & 21.94 & 0.85 & 0.77 \\
A2 & 357.911467 & -54.882801 & 21.49 & 0.81 & 0.69 \\
A3 & 357.906225 & -54.883464 & 22.30 & 0.84 & 0.74 \\
A4 & 357.907100 & -54.879967 & 21.46 & 0.91 & 0.77 \\
\enddata
\tablenotetext{a}{RA and Dec are epoch J2000.0 and are given in degrees.}
\tablenotetext{b}{$i$-band magnitudes for the knots are computed in 
     $3\arcsec$-diameter apertures, after first subtracting a model of the 
     BCG light derived using the Galfit galaxy fitting program \citep{peng02}.}
\tablenotetext{c}{$g-r$ and $r-i$ colors are computed from 
     $3\arcsec$-diameter SExtractor aperture magnitudes.}
\end{deluxetable}

\clearpage

\begin{deluxetable}{ccccccc}
\tabletypesize{\scriptsize}
\tablewidth{0pt}
\tablecaption{Cluster Galaxies\label{cluster_galaxies}}
\tablehead{
\colhead{Object ID\tablenotemark{a}} &
\colhead{RA\tablenotemark{b}} &
\colhead{Dec\tablenotemark{b}} &
\colhead{$i({\rm \tt MAG\_AUTO})$\tablenotemark{a}} &
\colhead{$g-r$\tablenotemark{a}} &
\colhead{$r-i$\tablenotemark{a}} &
\colhead{redshift $z$\tablenotemark{c}}}
\startdata
\sidehead{maxBCG Cluster Members\tablenotemark{d}}
15173 (BCG) &  357.908555 &  -54.881611 & $ 17.36 \pm 0.00 $ & $  1.86 \pm 0.01 $ & $  0.68 \pm 0.00 $ & $  0.3805 \pm 0.0003 $ \\
16097 &  357.972190 &  -54.856522 & $ 18.58 \pm 0.00 $ & $  1.87 \pm 0.02 $ & $  0.71 \pm 0.01 $ &  \\
16926 &  358.069064 &  -54.838013 & $ 18.67 \pm 0.01 $ & $  1.72 \pm 0.02 $ & $  0.75 \pm 0.01 $ &  \\
14954 &  357.990606 &  -54.881805 & $ 18.70 \pm 0.01 $ & $  1.78 \pm 0.02 $ & $  0.69 \pm 0.01 $ &  \\
14458 &  357.922935 &  -54.891348 & $ 19.05 \pm 0.00 $ & $  1.77 \pm 0.02 $ & $  0.66 \pm 0.01 $ & $  0.3844 \pm 0.0002 $ \\
15111 &  357.911305 &  -54.879770 & $ 19.10 \pm 0.01 $ & $  1.76 \pm 0.03 $ & $  0.64 \pm 0.01 $ &  \\
13772 &  357.854114 &  -54.908062 & $ 19.21 \pm 0.01 $ & $  1.76 \pm 0.02 $ & $  0.65 \pm 0.01 $ &  \\
14873 &  357.913389 &  -54.883120 & $ 19.22 \pm 0.01 $ & $  1.78 \pm 0.02 $ & $  0.66 \pm 0.01 $ &  \\
15204 &  357.917968 &  -54.874795 & $ 19.22 \pm 0.01 $ & $  1.81 \pm 0.02 $ & $  0.64 \pm 0.01 $ & $  0.3827 \pm 0.0002 $ \\
15305 &  357.929749 &  -54.874524 & $ 19.32 \pm 0.01 $ & $  1.66 \pm 0.02 $ & $  0.66 \pm 0.01 $ &  \\
15124 &  357.915316 &  -54.877257 & $ 19.39 \pm 0.01 $ & $  1.72 \pm 0.03 $ & $  0.65 \pm 0.01 $ & $  0.3929 \pm 0.0005 $ \\
11813 &  357.856326 &  -54.957697 & $ 19.40 \pm 0.01 $ & $  1.70 \pm 0.03 $ & $  0.63 \pm 0.01 $ &  \\
13629 &  357.781583 &  -54.911494 & $ 19.57 \pm 0.01 $ & $  1.67 \pm 0.03 $ & $  0.64 \pm 0.01 $ &  \\
16084 &  357.932492 &  -54.855191 & $ 19.62 \pm 0.01 $ & $  1.81 \pm 0.03 $ & $  0.67 \pm 0.01 $ & $  0.3864 \pm 0.0003 $ \\
14828 &  357.858498 &  -54.884039 & $ 19.69 \pm 0.01 $ & $  1.76 \pm 0.03 $ & $  0.67 \pm 0.01 $ &  \\
15056 &  357.929263 &  -54.878393 & $ 19.69 \pm 0.01 $ & $  1.83 \pm 0.03 $ & $  0.71 \pm 0.01 $ &  \\
13028 &  357.836914 &  -54.923702 & $ 19.69 \pm 0.01 $ & $  1.70 \pm 0.03 $ & $  0.65 \pm 0.01 $ &  \\
14267 &  357.742239 &  -54.897565 & $ 19.70 \pm 0.01 $ & $  1.66 \pm 0.04 $ & $  0.73 \pm 0.01 $ &  \\
13939 &  357.743518 &  -54.903441 & $ 19.72 \pm 0.01 $ & $  1.78 \pm 0.04 $ & $  0.70 \pm 0.01 $ &  \\
14892 &  357.917045 &  -54.881040 & $ 19.75 \pm 0.01 $ & $  1.67 \pm 0.03 $ & $  0.58 \pm 0.01 $ &  \\
17276 &  358.061857 &  -54.827190 & $ 19.83 \pm 0.01 $ & $  1.68 \pm 0.03 $ & $  0.67 \pm 0.01 $ &  \\
12997 &  357.992988 &  -54.925261 & $ 19.85 \pm 0.01 $ & $  1.79 \pm 0.04 $ & $  0.70 \pm 0.01 $ &  \\
14685 &  357.948912 &  -54.885316 & $ 19.85 \pm 0.01 $ & $  1.77 \pm 0.03 $ & $  0.70 \pm 0.01 $ &  \\
14727 &  357.914364 &  -54.884540 & $ 19.86 \pm 0.01 $ & $  1.74 \pm 0.04 $ & $  0.64 \pm 0.01 $ &  \\
12907 &  357.971817 &  -54.926860 & $ 19.88 \pm 0.01 $ & $  1.73 \pm 0.03 $ & $  0.71 \pm 0.01 $ &  \\
15525 &  357.891439 &  -54.867148 & $ 19.95 \pm 0.01 $ & $  1.85 \pm 0.04 $ & $  0.67 \pm 0.01 $ & $  0.3802 \pm 0.0005 $ \\
13874 &  357.767222 &  -54.904760 & $ 19.98 \pm 0.01 $ & $  1.76 \pm 0.04 $ & $  0.68 \pm 0.01 $ &  \\
14875 &  357.896375 &  -54.880676 & $ 20.00 \pm 0.01 $ & $  1.81 \pm 0.04 $ & $  0.67 \pm 0.01 $ &  \\
14827 &  357.956282 &  -54.883059 & $ 20.02 \pm 0.01 $ & $  1.73 \pm 0.04 $ & $  0.69 \pm 0.01 $ &  \\
14169 &  357.942454 &  -54.896963 & $ 20.05 \pm 0.01 $ & $  1.69 \pm 0.04 $ & $  0.67 \pm 0.01 $ &  \\
14620 &  357.906482 &  -54.885446 & $ 20.09 \pm 0.01 $ & $  1.65 \pm 0.03 $ & $  0.66 \pm 0.01 $ & $  0.3822 \pm 0.0004 $ \\
11254 &  357.792343 &  -54.968633 & $ 20.11 \pm 0.01 $ & $  1.76 \pm 0.03 $ & $  0.63 \pm 0.01 $ &  \\
19279 &  357.988291 &  -54.784591 & $ 20.12 \pm 0.01 $ & $  1.67 \pm 0.04 $ & $  0.67 \pm 0.01 $ &  \\
15027 &  357.880749 &  -54.878200 & $ 20.16 \pm 0.01 $ & $  1.76 \pm 0.04 $ & $  0.63 \pm 0.01 $ & $  0.3876 \pm 0.0006 $ \\
12805 &  357.947638 &  -54.929596 & $ 20.18 \pm 0.01 $ & $  1.70 \pm 0.04 $ & $  0.76 \pm 0.01 $ &  \\
13899 &  358.003818 &  -54.902294 & $ 20.19 \pm 0.01 $ & $  1.84 \pm 0.04 $ & $  0.66 \pm 0.01 $ &  \\
14741 &  357.943783 &  -54.884299 & $ 20.21 \pm 0.01 $ & $  1.85 \pm 0.05 $ & $  0.69 \pm 0.01 $ &  \\
12671 &  358.055413 &  -54.931583 & $ 20.22 \pm 0.01 $ & $  1.71 \pm 0.04 $ & $  0.59 \pm 0.01 $ &  \\
14843 &  357.901141 &  -54.880772 & $ 20.23 \pm 0.01 $ & $  1.72 \pm 0.04 $ & $  0.61 \pm 0.01 $ &  \\
14088 &  357.936003 &  -54.898050 & $ 20.27 \pm 0.01 $ & $  1.79 \pm 0.05 $ & $  0.69 \pm 0.01 $ & $  0.3816 \pm 0.0005 $ \\
14969 &  357.910388 &  -54.878452 & $ 20.29 \pm 0.01 $ & $  1.76 \pm 0.05 $ & $  0.64 \pm 0.01 $ &  \\
12875 &  357.935888 &  -54.927451 & $ 20.30 \pm 0.01 $ & $  1.80 \pm 0.04 $ & $  0.64 \pm 0.01 $ &  \\
13537 &  357.937414 &  -54.911304 & $ 20.31 \pm 0.01 $ & $  1.75 \pm 0.04 $ & $  0.68 \pm 0.01 $ & $  0.3849 \pm 0.0003 $ \\
15314 &  357.902668 &  -54.872019 & $ 20.34 \pm 0.01 $ & $  1.70 \pm 0.05 $ & $  0.60 \pm 0.01 $ & $  0.3841 \pm 0.0004 $ \\
14669 &  357.916522 &  -54.885626 & $ 20.36 \pm 0.01 $ & $  1.89 \pm 0.06 $ & $  0.73 \pm 0.02 $ & $  0.3862 \pm 0.0003 $ \\
14639 &  357.954384 &  -54.885672 & $ 20.36 \pm 0.01 $ & $  1.77 \pm 0.05 $ & $  0.67 \pm 0.01 $ &  \\
14232 &  357.904683 &  -54.895183 & $ 20.38 \pm 0.01 $ & $  1.76 \pm 0.04 $ & $  0.64 \pm 0.01 $ & $  0.3882 \pm 0.0003 $ \\
14703 &  357.865075 &  -54.884639 & $ 20.41 \pm 0.01 $ & $  1.81 \pm 0.05 $ & $  0.62 \pm 0.02 $ &  \\
14690 &  357.914016 &  -54.883857 & $ 20.42 \pm 0.01 $ & $  1.68 \pm 0.04 $ & $  0.64 \pm 0.01 $ &  \\
15463 &  357.882797 &  -54.868348 & $ 20.43 \pm 0.01 $ & $  1.71 \pm 0.05 $ & $  0.63 \pm 0.01 $ & $  0.3877 \pm 0.0004 $ \\
16005 &  357.991736 &  -54.856356 & $ 20.44 \pm 0.01 $ & $  1.70 \pm 0.05 $ & $  0.66 \pm 0.01 $ &  \\
15333 &  357.975175 &  -54.872090 & $ 20.44 \pm 0.01 $ & $  1.69 \pm 0.05 $ & $  0.63 \pm 0.01 $ &  \\
14972 &  357.909534 &  -54.878210 & $ 20.45 \pm 0.01 $ & $  1.87 \pm 0.06 $ & $  0.68 \pm 0.02 $ &  \\
14086 &  357.829155 &  -54.897455 & $ 20.48 \pm 0.01 $ & $  1.68 \pm 0.04 $ & $  0.59 \pm 0.01 $ &  \\
18418 &  357.768333 &  -54.801860 & $ 20.49 \pm 0.01 $ & $  1.65 \pm 0.04 $ & $  0.59 \pm 0.01 $ &  \\
13764 &  357.905657 &  -54.906287 & $ 20.50 \pm 0.01 $ & $  1.73 \pm 0.05 $ & $  0.65 \pm 0.01 $ &  \\
10692 &  357.819108 &  -54.981725 & $ 20.53 \pm 0.02 $ & $  1.71 \pm 0.06 $ & $  0.64 \pm 0.02 $ &  \\
15516 &  357.931958 &  -54.867137 & $ 20.56 \pm 0.01 $ & $  1.68 \pm 0.05 $ & $  0.61 \pm 0.01 $ & $  0.3785 \pm 0.0001 $ \\
19588 &  357.961069 &  -54.776725 & $ 20.57 \pm 0.02 $ & $  1.64 \pm 0.05 $ & $  0.64 \pm 0.02 $ &  \\
15002 &  357.897311 &  -54.877856 & $ 20.58 \pm 0.01 $ & $  1.70 \pm 0.05 $ & $  0.67 \pm 0.02 $ & $  0.3838 \pm 0.0004 $ \\
14800 &  357.901708 &  -54.880859 & $ 20.59 \pm 0.01 $ & $  1.73 \pm 0.05 $ & $  0.64 \pm 0.01 $ &  \\
15788 &  357.914426 &  -54.860804 & $ 20.61 \pm 0.01 $ & $  1.75 \pm 0.05 $ & $  0.64 \pm 0.02 $ & $  0.3821 \pm 0.0002 $ \\
15373 &  357.965957 &  -54.874282 & $ 20.61 \pm 0.01 $ & $  1.66 \pm 0.05 $ & $  0.64 \pm 0.02 $ & $  0.3856 \pm 0.0005 $ \\
13697 &  357.905543 &  -54.907479 & $ 20.64 \pm 0.01 $ & $  1.81 \pm 0.06 $ & $  0.68 \pm 0.02 $ & $  0.3782 \pm 0.0005 $ \\
15187 &  357.874035 &  -54.873868 & $ 20.65 \pm 0.01 $ & $  1.89 \pm 0.06 $ & $  0.64 \pm 0.02 $ & $  0.3868 \pm 0.0006 $ \\
18026 &  358.056024 &  -54.810258 & $ 20.66 \pm 0.01 $ & $  1.74 \pm 0.05 $ & $  0.70 \pm 0.02 $ &  \\
14378 &  357.875627 &  -54.892067 & $ 20.71 \pm 0.02 $ & $  1.66 \pm 0.05 $ & $  0.56 \pm 0.02 $ &  \\
14844 &  357.899766 &  -54.880469 & $ 20.72 \pm 0.04 $ & $  1.75 \pm 0.31 $ & $  0.72 \pm 0.09 $ &  \\
17455 &  357.997121 &  -54.822676 & $ 20.72 \pm 0.04 $ & $  1.54 \pm 0.40 $ & $  0.88 \pm 0.12 $ &  \\
17729 &  357.875358 &  -54.816995 & $ 20.74 \pm 0.01 $ & $  1.78 \pm 0.06 $ & $  0.64 \pm 0.02 $ &  \\
15068 &  358.094352 &  -54.876409 & $ 20.75 \pm 0.02 $ & $  1.64 \pm 0.07 $ & $  0.68 \pm 0.02 $ &  \\
15994 &  357.763211 &  -54.855887 & $ 20.82 \pm 0.02 $ & $  1.80 \pm 0.06 $ & $  0.64 \pm 0.02 $ &  \\
12892 &  358.014272 &  -54.926461 & $ 20.86 \pm 0.02 $ & $  1.65 \pm 0.07 $ & $  0.68 \pm 0.02 $ &  \\
15697 &  357.897819 &  -54.862968 & $ 20.86 \pm 0.02 $ & $  1.73 \pm 0.06 $ & $  0.55 \pm 0.02 $ & $  0.3789 \pm 0.0003 $ \\
12589 &  357.785585 &  -54.933352 & $ 20.87 \pm 0.03 $ & $  1.66 \pm 0.10 $ & $  0.63 \pm 0.03 $ &  \\
11976 &  357.893073 &  -54.951394 & $ 20.88 \pm 0.02 $ & $  1.62 \pm 0.06 $ & $  0.61 \pm 0.02 $ &  \\
14664 &  357.901167 &  -54.885034 & $ 20.89 \pm 0.02 $ & $  1.84 \pm 0.08 $ & $  0.68 \pm 0.02 $ &  \\
13901 &  357.824131 &  -54.902094 & $ 20.90 \pm 0.02 $ & $  1.69 \pm 0.06 $ & $  0.64 \pm 0.02 $ &  \\
14595 &  357.914211 &  -54.886290 & $ 20.93 \pm 0.03 $ & $  1.80 \pm 0.13 $ & $  0.66 \pm 0.03 $ &  \\
12863 &  357.874307 &  -54.926847 & $ 20.95 \pm 0.02 $ & $  1.66 \pm 0.07 $ & $  0.62 \pm 0.02 $ &  \\
15156 &  357.891357 &  -54.874854 & $ 20.95 \pm 0.03 $ & $  1.59 \pm 0.09 $ & $  0.56 \pm 0.03 $ &  \\
14825 &  357.922988 &  -54.880749 & $ 20.95 \pm 0.02 $ & $  1.63 \pm 0.07 $ & $  0.64 \pm 0.02 $ &  \\
12650 &  357.958048 &  -54.931820 & $ 20.96 \pm 0.02 $ & $  1.77 \pm 0.08 $ & $  0.64 \pm 0.02 $ &  \\
14407 &  357.917097 &  -54.891402 & $ 20.97 \pm 0.02 $ & $  1.75 \pm 0.07 $ & $  0.63 \pm 0.02 $ & $  0.3799 \pm 0.0004 $ \\
14939 &  357.872276 &  -54.878410 & $ 20.97 \pm 0.02 $ & $  1.75 \pm 0.07 $ & $  0.62 \pm 0.02 $ &  \\
14944 &  357.913836 &  -54.877816 & $ 20.98 \pm 0.03 $ & $  2.17 \pm 0.23 $ & $  0.64 \pm 0.05 $ &  \\
\sidehead{Other Spectroscopic Cluster Members\tablenotemark{e}}
14271 &  357.899268 &  -54.896523 & $ 19.13 \pm 0.01 $ & $  1.90 \pm 0.02 $ & $  0.71 \pm 0.01 $ & $  0.3814 \pm 0.0002 $ \\
15403 &  357.908847 &  -54.870362 & $ 19.52 \pm 0.01 $ & $  1.58 \pm 0.02 $ & $  0.67 \pm 0.01 $ & $  0.3860 \pm 0.0003 $ \\
15827 &  357.957606 &  -54.860595 & $ 19.64 \pm 0.01 $ & $  1.11 \pm 0.02 $ & $  0.45 \pm 0.01 $ & $  0.3900 \pm 0.0004 $ \\
15400 &  357.866255 &  -54.870713 & $ 19.69 \pm 0.01 $ & $  1.62 \pm 0.03 $ & $  0.65 \pm 0.01 $ & $  0.3768 \pm 0.0003 $ \\
14466 &  357.909595 &  -54.890780 & $ 20.47 \pm 0.02 $ & $  1.57 \pm 0.05 $ & $  0.60 \pm 0.02 $ & $  0.3827 \pm 0.0002 $ \\
14492 &  357.914762 &  -54.888447 & $ 20.88 \pm 0.02 $ & $  1.58 \pm 0.06 $ & $  0.64 \pm 0.02 $ & $  0.3899 \pm 0.0004 $ \\
13372 &  357.917399 &  -54.914403 & $ 20.98 \pm 0.02 $ & $  1.60 \pm 0.06 $ & $  0.65 \pm 0.02 $ & $  0.3803 \pm 0.0003 $ \\
14505 &  357.870029 &  -54.888283 & $ 21.27 \pm 0.02 $ & $  1.63 \pm 0.08 $ & $  0.66 \pm 0.03 $ & $  0.3860 \pm 0.0003 $ \\
\enddata
\tablenotetext{a}{Object ID numbers are from the SExtractor catalog obtained 
     using the $i$-band image for object detection.  The objects
     are ordered from bright to faint by $i$-band {\tt MAG\_AUTO}, starting
     with the BCG.
     $g-r$ and $r-i$ colors are computed from $3\arcsec$-diameter 
     aperture magnitudes.  The errors are simply statistical errors reported by
     SExtractor.  Not included are photometric calibration errors estimated
     to be 0.03-0.05 mag per filter.}
\tablenotetext{b}{RA and Dec are epoch J2000.0 and are given in degrees.}
\tablenotetext{c}{Redshifts measured from GMOS spectroscopy
                  (\S\ref{sec:redshifts}).}
\tablenotetext{d}{Galaxies, with $i < 21$, determined to be cluster members 
     using maxBCG color selection criteria.  Members are also limited to
     be within a radius 
     $r_{200}^{gal} = 1.51~h^{-1}~{\rm Mpc}$ ($= 6.88\arcmin$) of the BCG.
     See \S\ref{sec:cluster_properties} for details.}
\tablenotetext{e}{Additional galaxies determined to be cluster members via 
     GMOS spectroscopic redshifts (\S\ref{sec:redshifts}), but which did not 
     meet the maxBCG color selection criteria.}
\end{deluxetable}

\clearpage

\begin{deluxetable}{ccccccc}
\tabletypesize{\scriptsize}
\tablewidth{0pt}
\tablecaption{Other Galaxies\label{other_galaxies}\tablenotemark{a}}
\tablehead{
\colhead{Object ID\tablenotemark{b}} &
\colhead{RA\tablenotemark{c}} &
\colhead{Dec\tablenotemark{c}} &
\colhead{$i({\rm \tt MAG\_AUTO})$\tablenotemark{b}} &
\colhead{$g-r$\tablenotemark{b}} &
\colhead{$r-i$\tablenotemark{b}} &
\colhead{redshift $z$\tablenotemark{d}}}
\startdata
14193 &  357.895093 &  -54.901998 & $ 17.99 \pm 0.00 $ & $  1.59 \pm 0.01 $ & $  0.58 \pm 0.00 $ & $  0.2970 \pm 0.0003 $ \\
15313 &  357.893518 &  -54.875789 & $ 18.75 \pm 0.00 $ & $  1.11 \pm 0.01 $ & $  0.47 \pm 0.01 $ & $  0.2486 \pm 0.0002 $ \\
16682 &  357.903652 &  -54.840904 & $ 19.97 \pm 0.01 $ & $  1.40 \pm 0.03 $ & $  0.57 \pm 0.01 $ & $  0.3259 \pm 0.0002 $ \\
13520 &  357.887606 &  -54.911328 & $ 20.10 \pm 0.01 $ & $  0.51 \pm 0.01 $ & $  0.27 \pm 0.01 $ & $  0.0649 \pm 0.0001 $ \\
19352 & 357.902529 & -54.852090 & $20.18 \pm 0.01$ & $1.01 \pm 0.02$ & $0.25 \pm 0.01$ & $0.4214 \pm 0.0003$ \\
15509 &  357.941448 &  -54.869154 & $ 20.29 \pm 0.01 $ & $  1.57 \pm 0.05 $ & $  0.62 \pm 0.02 $ & $  0.4178 \pm 0.0005 $ \\
16409 &  357.890133 &  -54.846245 & $ 20.30 \pm 0.01 $ & $  0.89 \pm 0.02 $ & $  0.31 \pm 0.01 $ & $  0.3251 \pm 0.0002 $ \\
16570 &  357.911876 &  -54.843091 & $ 20.42 \pm 0.01 $ & $  0.77 \pm 0.03 $ & $  0.58 \pm 0.02 $ & $  0.1277 \pm 0.0002 $ \\
13423 &  357.960803 &  -54.913826 & $ 20.63 \pm 0.02 $ & $  1.01 \pm 0.03 $ & $  0.68 \pm 0.02 $ & $  0.2524 \pm 0.0002 $ \\
13620 &  357.889293 &  -54.909472 & $ 20.86 \pm 0.02 $ & $  1.66 \pm 0.08 $ & $  0.90 \pm 0.02 $ & $  0.5354 \pm 0.0004 $ \\
19257 & 357.902437 & -54.851578 & $21.45 \pm 0.03$ & $0.98 \pm 0.03$ & $-0.06 \pm 0.02$ & $0.2970 \pm 0.0004$ \\
16562 &  357.948746 &  -54.841891 & $ 21.58 \pm 0.03 $ & $  1.90 \pm 0.12 $ & $  0.61 \pm 0.03 $ & $  0.3595 \pm 0.0002 $ \\
\enddata
\tablenotetext{a}{Galaxies determined to be non-cluster members based on
     GMOS spectroscopic redshifts (\S\ref{sec:redshifts}).}
\tablenotetext{b}{Object ID numbers are from the SExtractor catalog obtained
     using the $i$-band image for object detection.  The objects
     are ordered from bright to faint by $i$-band {\tt MAG\_AUTO}.
     $g-r$ and $r-i$ colors are computed from $3\arcsec$-diameter
     aperture magnitudes.  The errors are simply statistical errors reported by
     SExtractor.  Not included are photometric calibration errors estimated
     to be 0.03-0.05 mag per filter.}
\tablenotetext{c}{RA and Dec are epoch J2000.0 and are given in degrees.}
\tablenotetext{d}{Redshifts measured from GMOS spectroscopy
                  (\S\ref{sec:redshifts}).}
\end{deluxetable}

\clearpage

\begin{deluxetable}{ccccccccll}
\tabletypesize{\scriptsize}
\tablewidth{0pt}
\tablecaption{NFW Fit Results\label{table_weak_lensing}}
\tablehead{
\colhead{Filter} & 
\colhead{$N$\tablenotemark{a}} &
\colhead{$m_1$\tablenotemark{b}}    &
\colhead{$m_2$\tablenotemark{b}}    &
\colhead{$(Q_{xx}+Q_{yy})_{min}$\tablenotemark{c}}    &
\colhead{$z_{crit}$\tablenotemark{d}}    &
\colhead{$M_{200} (10^{14} M_\sun)$\tablenotemark{e}}    &
\colhead{$c_{200}$\tablenotemark{e}}    &
\colhead{$\chi^2/{\rm dof}$\tablenotemark{f}} &
\colhead{$P$\tablenotemark{f}}}
\startdata
\sidehead{tangential shear}
$g$ & 1883 & 22.5 & 24.0 & 18.75 & 0.68 & $0.1^{+0.4}_{-0.1}$ & $> 45$ & 0.83 & 0.68 \\
$r$ & 7013 & 22.0 & 24.0 & 12.0 &  0.70 & $3.9^{+2.9}_{-2.1}$ & $6.5^{+5.3}_{-3.0}$ & 1.54 & 0.059 \\
$i$ & 3296 & 22.0 & 23.5 & 12.0 &  0.71 & $5.9^{+5.3}_{-3.8}$ & $3.7^{+13.1}_{-2.6}$ & 1.55 & 0.055 \\
$z$ & 2300 & 20.5 & 22.5 & 12.0 &  0.62 & $11.0^{+11.9}_{-7.1}$ & $1.8^{+3.6}_{-1.8}$ & 0.89 & 0.60 \\
\\
$i$+$r$ & 7995 & & & & 0.70 & $4.2^{+2.8}_{-2.1}$ & $6.1^{+4.9}_{-3.0}$ & 1.58 & 0.048 \\
$i$+$r$+$z$ & 8996 & & & & 0.70 & $5.0^{+2.9}_{-2.3}$ & $4.9^{+3.9}_{-2.2}$ & 1.48 & 0.077 \\
$i$+$r$+$z$+$g$ & 9424 & & & & 0.70 & $4.3^{+2.8}_{-2.2}$ & $5.2^{+5.4}_{-2.5}$ & 1.50 & 0.069 \\
$i$+$r$+$z$+SL(s)\tablenotemark{g} & 8996 & & & & 0.70 & $4.8^{+2.8}_{-2.2}$ & $6.2^{+3.2}_{-1.7}$ & 1.48 & 0.077 \\
$i$+$r$+$z$+SL(sg)\tablenotemark{g} & 8996 & & & & 0.70 & $4.9^{+2.9}_{-2.2}$ & $5.5^{+2.7}_{-1.6}$ & 1.48 & 0.077 \\
WL+SL+$\sigma_c$+$N_{200}$\tablenotemark{h} & 8996 & & & & 0.70 & $5.1^{+1.3}_{-1.3}$ & $5.4^{+1.4}_{-1.1}$ & 1.48 & 0.077 \\
\sidehead{B-mode shear}
$g$ & 1883 & 22.5 & 24.0 & 18.75 & 0.68 & $1.6^{+3.1}_{-1.5}$ & $6.5^{+10.1}_{-5.4}$ & 1.04 & 0.41 \\
$r$ & 7013 & 22.0 & 24.0 & 12.0 &  0.70 & $0.1^{+0.1}_{-0.1}$ & $> 63$ & 1.19 & 0.25 \\
$i$ & 3296 & 22.0 & 23.5 & 12.0 &  0.71 & $0.1^{+0.4}_{-0.1}$ & $> 0$ & 0.91 & 0.58 \\
$z$ & 2300 & 20.5 & 22.5 & 12.0 &  0.62 & $5.5^{+10.7}_{-5.2}$ & $0.3^{+1.2}_{-0.3}$ & 0.61 & 0.91 \\
\\
$i$+$r$ & 7995 & & & & 0.70 & $0.1^{+0.1}_{-0.1}$ & $> 51$ & 1.10 & 0.34 \\
$i$+$r$+$z$ & 8996 & & & & 0.70 & $0.1^{+0.1}_{-0.1}$ & $> 27$ & 0.80 & 0.71 \\
$i$+$r$+$z$+$g$ & 9424 & & & & 0.70 & $0.1^{+0.6}_{-0.1}$ & $> 0$ & 0.77 & 0.75 \\
\enddata
\tablenotetext{a}{Number of galaxies used in the weak lensing analysis in 
                  each filter.  For the multi-filter samples $N$ is the
                  number of {\em unique} galaxies.}
\tablenotetext{b}{SExtractor {\tt MAG\_AUTO} magnitude limits used to
                  define the galaxy sample.}
\tablenotetext{c}{Minimum Ellipto size $Q_{xx}+Q_{yy}$ used to
                  define the galaxy sample.}
\tablenotetext{d}{The source redshift at which $1 / \Sigma_{crit}$ is the same
                  as the effective value computed by integration over the
                  source galaxy redshift distribution, as described in 
                  \S\ref{sec:shear_profiles}.}
\tablenotetext{e}{Best-fit NFW profile parameters: mass $M_{200}$ and 
                  concentration $c_{200}$. Errors are 1-parameter, 
                  $1\sigma$ values, as determined by where $\Delta \chi^2 = 1$.
                  The uncertainties on $M_{200}$ are rounded off to 
                  the nearest $0.1 \times 10^{14} M_\sun$.
                  Note that for most of the cases (primarily B-mode fits)
                  where there is no significant mass detection, 
                  we provide only a $1\sigma$ lower limit on $c_{200}$, 
                  which is otherwise not constrained on the high side 
                  even at $1\sigma$, up to the upper bound value 
                  $c_{200} = 10^4$ that we have checked.  
                  Joint 2-parameter error contours for
                  select samples are shown in Fig.~\ref{contours_M200_c}.}
\tablenotetext{f}{$\chi^2$ per degree of freedom (dof) relative to a null 
                  hypothesis of zero shear.  (This is {\em not} the 
                  $\chi^2/{\rm dof}$ of the NFW fit, which is very close to
                  one in all cases.)  $P$ is the probability of
                  exceeding the observed $\chi^2/{\rm dof}$.  The number of 
                  degrees of freedom for this $\chi^2$ test is always 20, 
                  i.e., the number of radial bins plotted in 
                  Figures~\ref{nfw_fit_ir}-\ref{nfw_fit_irz_sl}.}
\tablenotetext{g}{Fit results derived from combined weak plus strong 
                  lensing (``SL'') constraints. ``(s)'' denotes the case 
                  where we estimated the dark matter mass within the Einstein 
                  radius by subtracting off just a stellar mass contribution, 
                  while ``(sg)'' is the case where we also subtracted off an 
                  estimated gas mass contribution.  See \S\ref{sec:combined}
                  for details.}
\tablenotetext{h}{Fit results derived from combined weak lensing ($i+r+z$),
                  strong lensing (SL(sg)), cluster velocity dispersion
                  ($\sigma_c$), and cluster richness ($N_{200}$) constraints. 
                  See \S\ref{sec:all_combined} for details.}
\end{deluxetable}

\clearpage

\begin{deluxetable}{lccccc}
\tabletypesize{\scriptsize}
\rotate
\tablewidth{0pt}
\tablecaption{Source Galaxy Star Formation Rates\tablenotemark{a}\label{table_sfr}}
\tablehead{
\colhead{Knot} & 
\colhead{$f(\nu)_{[OII]}\ ({\rm erg\ s^{-1}\ cm^{-2}})$} &
\colhead{$f(\nu)_L\ ({\rm erg\ s^{-1}\ cm^{-2}})$} &
\colhead{$f(\nu)_S\ ({\rm erg\ s^{-1}\ cm^{-2} Hz^{-1}})$} &
\colhead{SFR (($f_{lens}=39$) $M_\sun \ {\rm yr}^{-1}$)} &
\colhead{SFR (($f_{lens}=141$) $M_\sun \ {\rm yr}^{-1}$)}}
\startdata
A1 & $1.06\pm 0.04\times 10^{-15}$ & $1.71\pm 0.06\times 10^{-16}$ &
$1.36\pm 0.02\times 10^{-28}$ & $3.9\pm 1.1$ & $1.1\pm 0.4$ \\
A2 & $0.84\pm 0.04\times 10^{-15}$ & $1.43\pm 0.06\times 10^{-16}$ &
$1.43\pm 0.02\times 10^{-28}$ & $3.1\pm 0.9$ & $0.85\pm 0.4$\\
A3 & $2.09\pm 0.10 \times 10^{-15}$ & $1.28\pm 0.06\times 10^{-16}$ &
$0.51 \pm 0.01\times 10^{-28}$ & $7.7\pm 2.2$ & $2.1\pm 0.4$\\
A4 & $1.02\pm 0.02\times 10^{-15}$ & $2.83\pm 0.06\times 10^{-16}$ &
$2.33\pm 0.02\times 10^{-28}$ & $3.7\pm 1.1$ & $1.0\pm 0.4$\\
\enddata
\tablenotetext{a}{See \S\ref{sec:SFR} for the definitions of the various
                  fluxes $f(\nu)$. Fluxes quoted are measured values.  $f_{lens}$ is the lens magnification.}
\end{deluxetable}




\end{document}